\documentclass[aps,12pt,axodraw,nofootinbib,superscriptaddress,aps]{revtex4}

\pdfoutput=1
\usepackage{epsfig}
\usepackage{amsmath}
\usepackage{bm}
\usepackage{times}
\usepackage{graphicx}
\usepackage{epstopdf}
\usepackage{amsfonts}
\usepackage{bm}
\usepackage{epsfig}
\usepackage{graphics}
\usepackage{xspace}
\usepackage[usenames]{color}

\usepackage{amstext}
\usepackage{amssymb}

\def\bea{\begin{eqnarray}}
\def\eea{\end{eqnarray}}
\def\ba{\begin{eqnarray}}
\def\ea{\end{eqnarray}}
\def\be{\begin{equation}}
\def\ee{\end{equation}}

\def\beq{\begin{equation}}
\def\eeq{\end{equation}}

\begin{document}

\title{\Large Limits on Anomalous Couplings of the Higgs to Electroweak Gauge Bosons from LEP and LHC}

\author{Eduard Mass\'o}
\affiliation{Departament de F\'{\i}sica and Institut de F\'{\i}sica d'Altes Energies,
Universitat Aut\`onoma de Barcelona, Barcelona, Spain}
\author{Ver\'onica Sanz}
\affiliation{Theory Division, Physics Department, CERN, CH-1211 Geneva 23, Switzerland}
\affiliation{Department of Physics and Astronomy, York University, Toronto, ON, Canada, M3J 1P3}

\begin{abstract}
In this paper we assume the Higgs is an elementary scalar, and study how new physics could affect  its couplings to electroweak gauge bosons. Adding LHC data to LEP data provides new, more stringent limits, particularly when the Higgs to two photon decay signal strength is taken into account. We then study the effect of anomalous angular correlations in the decay to WW*. We obtain a new limit on the rare decay to photon-Z, and use it to constrain Supersymmetry, to find that staus with large mixing would be most sensitive to this channel.  We also use these limits to constrain radion exchange in Warped Extra-Dimensions, finding a limit on the radion mass and interaction scale of the order of TeV. Finally, we have extrapolated the current data to obtain prospects for the full 2012 dataset. 
\end{abstract}
%%%%%%%%%%%%%%%%%%%%%%%%%%%%%%%%%%%%%%%%%%%%%%%%%%%%%%%%%%%%%%%%%%%%
\maketitle
%%%%%%%%%%%%%%%%%%%%%%%
\section{Introduction}\label{intro}
%%%%%%%%%%%%%%%%%%%%%%%

The new particle recently observed at around 125 GeV at LHC \cite{:2012gk} has properties consistent with  the Standard Model Higgs boson.
More precise measurements of its couplings will provide detailed information
 on the mechanism of Electroweak
Symmetry Breaking (EWSB) of the Standard Model (SM). An impressive effort in the community is made to test the nature of the resonance~\cite{fits}, whether is composite or fundamental, or even testing its scalar properties~\cite{HQN}. 

In this paper we assume that the discovered resonance is a {\it fundamental } scalar, and (at least partly) responsible for EWSB. Our objective is to show how LEP indirect data, and now LHC direct Higgs measurements are shaping our understanding of new physics from the so-far only resonance uncovered at the LHC, which we call "the Higgs" below. 

We now proceed to set the notation used in the paper, from the Higgs sector to the basis of effective operators adopted  thorough paper.

In the SM, the Higgs particle $H$ forms part of a $SU(2)_L$ doublet,
\begin{equation}
\Phi = \frac{1}{\sqrt{2}}
\left( 
\begin{array}{c} 
\phi_1+ i \phi_2 \\ 
v + H + i \phi_3 
\end{array} 
\right)
\end{equation}
where $v$ is the vacuum expectation value of the neutral part of the doublet. 
The term in the Lagrangian responsible for the gauge bosons masses is 
\begin{equation}
\label{ }
 (D_\mu \Phi)^\dagger (D^\mu \Phi)
\end{equation}
Here the covariant derivative is given by
\begin{equation}
D_\mu \Phi = \left( \partial_\mu + i \frac{g}{2} \tau^a W^a_\mu + i \frac{g^\prime }{2}  B_\mu \right) \Phi,
\end{equation}
where $\tau^a$ are the Pauli matrices, $g$ and $g^\prime$ are the $SU(2)_L$ and $U(1)_Y$ gauge couplings respectively, 
and the corresponding gauge fields $V$ are $W^a_\mu$ and $B_\mu$.

EWSB gives rise at the same time to the generation of the weak gauge boson masses and to their couplings $HVV$,
\begin{equation}
\label{SM}
\frac{g^2 v^2}{4} \,  W^+_\mu W^{- \, \mu} \left( 1 + 2 \frac{H}{v} \right)
+ \frac{1}{2} \, \frac{(g^2 + g^{\prime\, 2} )v^2}{4} \, Z_\mu Z^\mu \left( 1 + 2 \frac{H}{v} \right)
\end{equation}
where we have defined the fields $W^{\pm} = (W^1 \mp i W^2)/\sqrt{2}$. In eq.(\ref{SM}), valid at tree-level,
we appreciate the presence of the custodial symmetry: in the limit $g^\prime \rightarrow 0$, the masses of the three weak gauge bosons are equal,
and the couplings to the Higgs are identical.

Since the EWSB mechanism is at the origin of the mass of $W$ and $Z$ and at the same time determines the couplings of the Higgs to them,
the study of the couplings of the Higgs to electroweak bosons is expected to shed light on the EWSB mechanism.
With this in mind, in this paper we deal with the coupling of the Higgs to electroweak bosons.
Along the same philosophy, we are not considering anomalous coupling of the Higgs with the gluon $G$. This is a simplification, but also motivated by the tight direct bounds on strong production from the LHC. New physics able to modify the $H G G$ coupling would be necessarily charged under $SU(3)_c$. An example would be a new heavy color triplet, such as the stop in Supersymmetry. This colored particle would have an important effect in the Higgs  couplings to both gluons and photons, and their effect is very correlated such that the stop effect can be rephrased in terms of higher order operators~\cite{NSUSY}.  

As we said, in this paper we assume that the particle observed at LHC at 125 GeV is the fundamental Higgs particle $H$, but we also suppose that there is New Physics at higher energies
which might induce some relatively small changes in the $H$ properties. Specifically, we will concentrate on the $HVV$ couplings which are not of the form shown
in (\ref{SM}). We will also assume that the effects beyond the SM can be described in terms of effective Lagrangians valid up to a high-energy scale $\Lambda$. 

How heavy has to be the New Physics to fall into this analysis, based on effective operators? All processes concerning the Higgs at the LHC have relatively small energy exchange. For example, in the dominating production mechanism, gluon fusion, one can write the factorization
\bea
\sigma_{prod} (g g \to H) \simeq \frac{\pi^2}{8 \, m_H} \, \Gamma( H \to gg) \, \delta (\hat{s}-m_H^2)
\eea
which is just based on the narrow width approximation~\cite{Sally-NWA}. Therefore, in Higgs production through gluon fusion, the typical exchange in momentum is $Q^2 \simeq m_H^2$, and New Physics at a scale $\Lambda \gg m_H$ would lead to a sensible effective theory. The same reasoning follows for other production mechanisms, such as vector bosons fusion or associated production.

%%%%%%%%%%%%%%%%%%%%%%%
\section{The language of effective operators}\label{operators}
%%%%%%%%%%%%%%%%%%%%%%%

Let us consider the effective Lagrangian 
\bea
{\cal L}_{eff} = \sum_i \frac{f_i}{\Lambda^2} {\cal O}_i
\label{Leff}
\eea
with the subset of operators ${\cal O}_i$ which modify the $HVV$ vertices~\footnote{For bounds on other operators not involving the Higgs see Ref.~\cite{Paco}.}. These operators  have been investigated by 
\cite{HagiwaraPRD,DeRujula, Hagiwara:1992eh} before the LHC discovery, and by \cite{Corbett:2012dm} after the Higgs discovery.
The operators ${\cal O}_i$ are dimension-six and thus they are suppressed by a high-energy scale $\Lambda$. Throughout our paper we use the convention and notations which were used in  \cite{Hagiwara:1992eh,Corbett:2012dm}.
 
The list of operators is not long. We start with operators containing the scalar field $\Phi$ and its derivative.
There is a first operator which breaks custodial symmetry at tree-level,
\begin{equation}
\label{ophi1}
{\cal O}_{\Phi,1} =   (D_\mu\Phi)^\dagger \Phi \ \ 
 \Phi^\dagger (D^\mu \Phi)
\end{equation}
and there are two which preserve it~\footnote{Note that these two operators are not independent, as they are related by a non-linear field redefinition, $\Phi\to \Phi (1+ \alpha \,  \Phi^{\dagger} \Phi$), with a suitable parameter $\alpha$; see Ref.~\cite{SILH}.},
\begin{eqnarray}
\label{ophi2}
{\cal O}_{\Phi,2} & = & \frac{1}{2}  \  \partial_\mu(\Phi^\dagger \Phi) \
 \partial^\mu(\Phi^\dagger \Phi) \\
\label{ophi4}
{\cal O}_{\Phi,4} & = & (\Phi^\dagger \Phi)\ \
(D_\mu \Phi)^\dagger (D^\mu \Phi)
\end{eqnarray}

The list continues with five operators involving the scalar field and the field strengths
\begin{eqnarray}
\label{bmunu}
B_{\mu\nu} & = & \partial_\mu B_\nu - \partial_\nu B_\mu  \\
\label{wmunu}
W^a_{\mu\nu} & = & \partial_\mu W^a_\nu - \partial_\nu W^a_\mu - g\epsilon^{abc}
W_\mu^b W_\nu^c 
\end{eqnarray}
In this paper, we use the re-scaled field stregths $\widehat   B_{\mu\nu} = i (g^\prime /2) B_{\mu\nu}$ and $\widehat   W_{\mu\nu} = i (g/2) \tau^a W^a_{\mu\nu}$.

There is a first operator that contributes at tree-level to the $B-W^3$ mixing,
\begin{equation}
\label{obw}
{\cal O}_{BW}  =  
   \Phi^\dagger \widehat W^{\mu\nu}  \Phi  \ \widehat   B_{\mu\nu} 
\end{equation}
and four other operators
\begin{eqnarray}
\label{ow}
{\cal O}_{W} & = & (D_\mu\Phi)^\dagger 
                 \widehat  W^{\mu\nu}(D_\nu\Phi) \\
\label{ob}
{\cal O}_{B} & = & 
(D_\mu\Phi)^\dagger 
     (D_\nu\Phi) \  \widehat  B^{\mu\nu} \\
\label{oww}
{\cal O}_{WW} & = & \Phi^\dagger \widehat  W^{\mu\nu} \widehat W_{\mu\nu} \Phi 
 = -\frac{g^2}{4}
(\Phi^\dagger \Phi)
W^{a\,\mu\nu}W^a_{\mu\nu}  \\
\label{obb}
{\cal O}_{BB} & = & (\Phi^\dagger \Phi) \
\widehat   B^{\mu\nu} \widehat  B_{\mu\nu}  
\end{eqnarray}

Let us now discuss which operators we will consider in our paper.
First, the two operators (\ref{ophi1}) and (\ref{obw}) have tree level effect on precision electroweak observables and therefore are subject to very strict constraints. Due to this reason, we will not consider them: the LHC is not providing more information on those operators. Second, the operators (\ref{ophi2}) and (\ref{ophi4}) affect the $HVV$ couplings through a Higgs field renormalization, so that the induced effect have the same form than in (\ref{SM}), namely  it is of the form $H W^\mu W_\mu$ and $H Z^\mu Z_\mu$. 
Such normalization effects will be rather hard to extract in the near future~\footnote{See Ref.~\cite{Bonnet} for a study of the LHC limits on these operators.}, so we will not consider these two operators here. In summary, in this paper we will work out the consequences of the four operators ${\cal O}_{W}, {\cal O}_{B},{\cal O}_{WW},{\cal O}_{BB}$, contributing to ${\cal L}_{eff} $ in (\ref{Leff}) when added to the SM Lagrangian, with a total Lagrangian
\begin{equation}
\label{Ltotal}
{\cal L}= {\cal L}_{SM}  +{\cal L}_{eff}  \ .
\end{equation}

%%%%%%%%%%%%%%%%%%%%%%%
\section{Constraints from Precision Electroweak Physics}\label{LEP}
%%%%%%%%%%%%%%%%%%%%%%%

The four operators (\ref{ow}) - (\ref{obb}) contribute to precision electroweak observables measured for example at LEP and at Tevatron.
The observed experimental values severely constrain the presence of those operators. 

Let us start with data coming from measurements of $Z$-pole observables, $W$-mass, and low energy experiments. The standard way to proceed is to use the $S,T$, and $U$ parameters to find the bounds\footnote{We shall use the definitions of these parameters in the PDG \cite{PDG}.}. The operators we are considering contribute at the one-loop level to $S,T$, and $U$. The corresponding expressions have been calculated in \cite{Alam:1997nk}.
They read
\bea
\alpha S  =   \frac{ { e}^2}{96 \pi^2} & \Bigg\{ &
3  \left[  \epsilon_W + \epsilon_B \right]  \ \frac{m_H^2}{v^2} \log  \frac{\Lambda^2}{m_H^2} \nonumber  \\
& & + \ 2   \left[ (5 { c}^2 - 2) \epsilon_W - (5 { c}^2 - 3) \epsilon_B \right] \ \frac{m_Z^2}{v^2} \log  \frac{\Lambda^2}{m_H^2} \nonumber  \\
& & - \ \left[ (22 { c}^2 - 1) \epsilon_W - (30 { c}^2 + 1) \epsilon_B \right]  \ \frac{m_Z^2}{v^2} \log  \frac{\Lambda^2}{m_Z^2}  \nonumber\\
& & - \ 24 \left[ { c}^2 \epsilon_{WW} + { s}^2 \epsilon_{BB} \right] \ \frac{m_Z^2}{v^2} \log  \frac{\Lambda^2}{m_H^2} 
\ \Bigg\}
\label{expressionS}
\eea

\bea
\alpha T  =   \frac{ 3 { e}^2}{64 \pi^2 }\  \frac{1}{{ c}^2} & \Bigg\{ &  
  \epsilon_B \ \frac{m_H^2}{v^2} \log  \frac{\Lambda^2}{m_H^2} \nonumber  \\ 
& & + \  \left[ { c}^2 \epsilon_W + \epsilon_B  \right] \ \frac{m_Z^2}{v^2} \log  \frac{\Lambda^2}{m_H^2} \nonumber  \\
& & + \   \left[  2 { c}^2 \epsilon_W + ( 3 { c}^2 -1) \epsilon_B \right] \ \frac{m_Z^2}{v^2} \log  \frac{\Lambda^2}{m_Z^2}  
\ \Bigg\}
\label{expressionT}
\eea

\bea
\alpha U  =   \frac{ { e}^2 }{48 \pi^2 } \ { s}^2 \ & \Bigg\{ &  
\left[ 4  \epsilon_W  - 5 \epsilon_B  \right]  \ \frac{m_Z^2}{v^2} \log  \frac{\Lambda^2}{m_H^2} \nonumber  \\ 
& & + \ \left[ - 2  \epsilon_W  + 5 \epsilon_B \right] \ \frac{m_Z^2}{v^2} \log  \frac{\Lambda^2}{m_Z^2}
 \ \Bigg\}
 \label{expressionU}
\eea
Here we have defined
\begin{equation}
\label{}
  \epsilon_i = f_i \, \frac{v^2}{\Lambda^2}
\end{equation}
for $i=W,B,WW,BB$. The quantities $e$,  $c= \cos \theta_W$ and $s= \sin \theta_W$ are $\overline{MS}$ couplings. To have limits on the $\epsilon_i$ 
using $S,T,U$ in (\ref{expressionS}-\ref{expressionU}) we need to
specify the value of $\Lambda$ in the logarithm; we shall use $\Lambda=1$ TeV. 
Changing the value of $\Lambda$ does not change
very much the limits we shall find on the different $\epsilon_i$; in fact, as we increase $\Lambda$, the limits on $\epsilon_i$ tighten.  
The presence of the logarithm is because our operators affect $S,T,U$ at a one-loop order. 
As we will see, the induced $\epsilon_i$ for observables at LHC are at tree-level and do not have such logarithm of $\Lambda$.

The adimensional parameters $\epsilon_i$ encode the strength coefficient $f_i$ of the operators as well as the ratio
among the Fermi scale $v$ and the New Physics scale $\Lambda$. An alternative, which is also used in the current literature,
would be to specify the scale $\Lambda$ and show the limits on $f_i$. If one chooses $\Lambda=1$ TeV, any limit on a specific $\epsilon_i$
shown in our paper translates into a limit on the corresponding $f_i$ given by $f_i \simeq 16 \epsilon_i$.

We stress that the loops contributing to $S,T$, and $U$, as well as to the other observables contain quadratic divergences which cancel
because we are using an effective Lagrangian that contains gauge-invariant operators~\cite{DeRujula,HagiwaraPRD}. Only the logarithmic terms
remain; we have kept only this logarithmic part of the calculation and not the constant terms.

Let us now move to the physics of the triple gauge-boson vertices $V^3$, since
the operators ${\cal O}_{W}$ and ${\cal O}_{B}$  have contributions at tree level, and therefore
LEP2 measurements can constrain them.
In the presence of these two operators the relevant part of the total Lagrangian (\ref{Ltotal}) contains 
three parameters, $\Delta g_1^Z$, $\Delta \kappa_Z$ and $\Delta \kappa_\gamma$
\begin{eqnarray}
{\cal L}_{V^3}  &=& 
  i   \frac{ec}{s} \,  [  
(1+ \Delta g_1^Z) \, ( W^-_{\mu\nu} W^{+ \, \mu} 
  - W^+_{\mu\nu} W^-_{\mu} ) Z^{\nu}  
 +  (1+ \Delta \kappa_Z) \, W_\mu^- W_\nu^+ Z^{\mu\nu}
 ] \nonumber \\
&& + i e   \, [  
\, ( W^-_{\mu\nu} W^{+ \, \mu} 
  - W^+_{\mu\nu} W^-_{\mu} ) A^{\nu} 
  +  (1+ \Delta \kappa_\gamma) \, W_\mu^- W_\nu^+ A^{\mu\nu}
 ] 
\label{WWV}
\end{eqnarray}
where we have defined the field strengths corresponding to the abelian part,
\begin{equation}
\label{ }
V^{\mu \nu} = \partial^\mu V^\nu - \partial^\nu V^\mu
\end{equation}
with $V=\gamma, W,Z$. 

The deviations from the standard model due to the new parameters are given by \cite{HagiwaraPRD}
\begin{eqnarray}
\Delta g_1^Z & = &  \frac{e^2}{8s^2 c^2} \, \epsilon_W  \label{g1Z} \\
\Delta \kappa_Z & = &  \frac{e^2}{8s^2 c^2} \, ( c^2 \epsilon_W - s^2 \epsilon_B)  \label{kZ} \\
\Delta \kappa_\gamma & = &  \frac{e^2}{8s^2 } \, (  \epsilon_W + \epsilon_B)  \label{kg}
\end{eqnarray}

\subsection{Limits in the one parameter space}

To bound each operator we first use the recent limits found by Erler \cite{Erler:2012wz}
on the electroweak parameters $S,T$ and $U$.
The result of the fit to electrowek data for $M_H=125$ GeV is 
\begin{eqnarray} 
S  &=& 0.00 \pm 0.10 \nonumber \\
T  &=& 0.02 \pm 0.11 \nonumber \\
U  &=& 0.04 \pm 0.09
\label{STUbounds}
\end{eqnarray}

Imposing these experimental limits to each operator separately, i.e., not allowing for
cancellations between differents operators, and working at 95\%CL, we obtain
\begin{eqnarray}
-1.9  \,  \leqslant  & \epsilon_W & \leqslant  \, 2.3 \label{LEPbounds2A} \\
-0.90  \,  \leqslant  & \epsilon_B & \leqslant  \,  0.90 \\
-1.5  \,  \leqslant  & \epsilon_{WW} & \leqslant \,  1.5 \\
-5.6  \,  \leqslant  & \epsilon_{BB} & \leqslant  \,  5.6
\label{LEPbounds2}
\end{eqnarray}
Actually, it is $S$ and $T$ which restrict the $\epsilon_i$ parameters, with $U$ not playing a role. Indeed, one expects the new physics effect on $U$ to be suppressed
\cite{Peskin:1991sw} by $v^2/\Lambda^2$ respect to $T$.

Now we would like to find the analogous bounds coming from $V^3$ data.
We use the LEP2 experimental limits on $\Delta g_1^Z, \Delta \kappa_Z$ and $\Delta \kappa_\gamma$  as compiled by the PDG \cite{PDG},
\begin{eqnarray}
1+\Delta g_1^Z & = &  0.984^{+0.022}_{-0.019} \label{expg1Z} \\
1+ \Delta \kappa_Z & = &  0.924^{+0.067}_{-0.061}  \label{expkZ} \\
1 + \Delta \kappa_\gamma & = & 0.973^{+0.044}_{-0.045} \label{expkg}
\end{eqnarray}
We should stress that each of these bounds is obtained setting the other two parameters to their SM values, i.e. equal to zero.
Strictly speaking, the way these limits are extracted do not lead rigorously to individual bounds on $\epsilon_W$ and $\epsilon_B$.
For example, if we wish to get a limit on $\epsilon_W$ with  $\epsilon_B=0$ , we should allow 
the three parameters in (\ref{g1Z})-(\ref{kg}) to be non zero, in the proportions indicated by the equations and then compare with
experiment. 

However, there is a way out to this problem. Since the experimental bound on $\Delta g_1^Z$ leads  to the tightest  constraint,
and this depends only on $\epsilon_W$,
it is a very good approximation to neglect $\epsilon_W$ in (\ref{kZ}) and (\ref{kg}). Actually, when we make this approximation, we realize
that (\ref{expkg}) is more constraining than (\ref{expkZ}), so we can use only the limit coming from (\ref{expkg}).
With these approximations, we can get bounds on the strength
of the effective operators. Working at 95\%CL, we obtain
\begin{eqnarray}
-0.73 \, \leqslant  & \epsilon_W & \leqslant 0.38 \, \label{LEP2boundsA}  \\
-2.1 \,   \leqslant  & \epsilon_B & \leqslant  \, 1.04
\label{LEP2bounds}
\end{eqnarray}

We finally stress that LHC is also providing experimental data \cite{ATLAS:2012mec}
  on the $V^3$ vertices which in the future may be competitive with the old LEP2 data we use here.
  
%%%%%%%%%%%%%%%%%%%%%%%%%%
\subsection{Limits on the two parameter space}
%%%%%%%%%%%%%%%%%%%%%%%%%%

Although we do not expect fine tuned cancellations among the contributions of different operators,
it is interesting to see what happens when we consider two operators at the same time. Since the structure
of the operators ${\cal O}_{W}$ and ${\cal O}_{B}$ is very similar we shall consider the situation where they
are both generated by new physics existing at higher energy scales and investigate the bounds coming
from precision electroweak physics, including LEP2. For the same reasons we shall also consider the case where we
have the simultaneous effects of ${\cal O}_{WW}$ and ${\cal O}_{BB}$. In Secs.~\ref{gHAZsec} and ~\ref{toy}, 
we will provide examples of this common generation in the case of Supersymmetry and Extra-Dimensions.

For this exercise we shall use the LEP2 data, eqs. (\ref{expg1Z}) and (\ref{expkg}),  as well as
the limits on $S$ and $T$ when $U=0$ is fixed \cite{Erler}
\begin{eqnarray} 
S  &=& 0.02 \pm 0.08 \nonumber \\
T  &=& 0.05 \pm 0.07 
\label{STbounds}
\end{eqnarray}
with a correlation coefficient of 0.89. These values
are obtained\footnote{We thank Jens Erler for kindly providing us with these limits.} using the same data input as in \cite{Erler:2012wz}.

\begin{figure}[h!]
\centering
\includegraphics[scale=0.8]{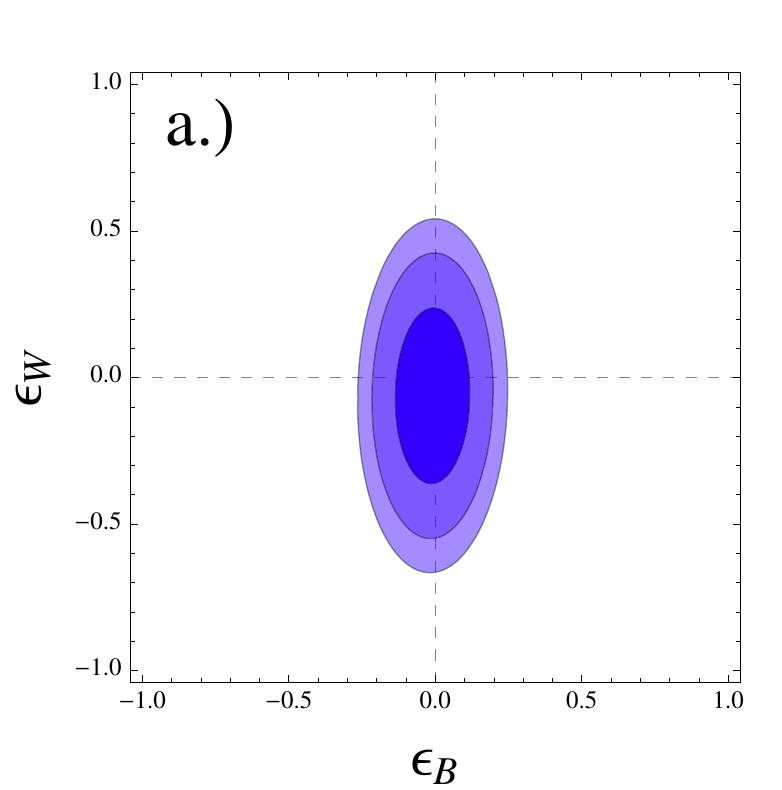} \hspace*{1cm}
\includegraphics[scale=0.8]{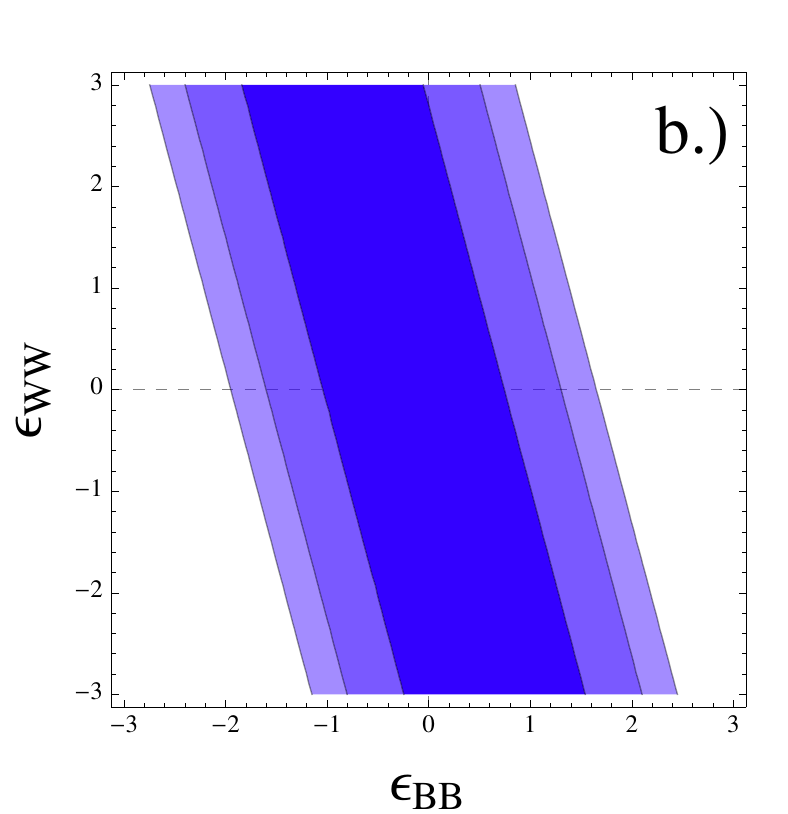}
\caption{\it The 68\%, 95\% and 99\% CL allowed regions in the parameters $(\epsilon_W,\epsilon_B)$ and  $(\epsilon_{WW},\epsilon_{BB})$. We use the
limits (\ref{STbounds} with the U parameter fixed to zero as well as (\ref{expg1Z}) and (\ref{expkg}).}
\label{STfigure}
\end{figure}

Figure \ref{STfigure} shows the 68\%, 95\% and 99\% confidence level (CL) allowed regions in the $(\epsilon_W,\epsilon_B)$ and  $(\epsilon_{WW},\epsilon_{BB})$ planes.
As expected, the contours in $(\epsilon_W,\epsilon_B)$ are ellipses. However, the contours in the plane $(\epsilon_{WW},\epsilon_{BB})$ are stripes because the
corresponding operators on the one hand do no modify the $V^3$ vertices, and on the other hand they are custodial preserving, 
and thus there is a single constraint coming from the $S$ parameter.

%%%%%%%%%%%%%%%%%%%%%%%
\section{LHC Bounds}\label{LHC}
%%%%%%%%%%%%%%%%%%%%%%%

In this Section we shall study the constraints on the operators (\ref{ow}) - (\ref{obb}) that can be obtained by analyzing Higgs {\it decays} at LHC. We do not consider the effect on the production, although the associated production channels ($q \bar{q} \to V^* \to V + H$) and the vector boson fusion ($q q' \to q q' V V \to q q' H$) are sensitive to the operators considered in this paper. The reason is that the production is largely dominated by gluon fusion processes, hence the decay rates are the best way to bound the operators.

\subsection{The translation between effective operators and Higgs couplings}

The contributions of our operators to the $HWW$ and $HZZ$ vertices have a different form that the SM expression (\ref{SM}). One can write
\begin{eqnarray}
\nonumber
\Delta {\cal L}_{HZZ}  & = & g^{(1)}_{HZZ} \ Z_{\mu \nu} Z^{\mu} \partial^\nu H \ + \
g^{(2)}_{HZZ} \ Z_{\mu \nu} Z^{\mu \nu} H \\
\Delta {\cal L}_{HWW}  & = & g^{(1)}_{HWW} \ \left( W^+_{\mu \nu} W^{- \mu} \partial^\nu H   + {\rm h.c.}\right) \ + \
g^{(2)}_{HWW} \ W^+_{\mu \nu} W^{- \mu \nu} H 
\label{LHVV}
\end{eqnarray}
where we have defined the field strengths corresponding to the abelian part,
\begin{equation}
\label{ }
V^{\mu \nu} = \partial^\mu V^\nu - \partial^\nu V^\mu
\end{equation}

The couplings in (\ref{LHVV}) are easily obtained \cite{HagiwaraPRD,Hagiwara:1992eh,Corbett:2012dm},
\begin{eqnarray}
\nonumber 
g^{(1)}_{HZZ}  & = &  \frac{e^2}{4v} \,  \left(  \frac{1}{c^2}  \epsilon_B \, +\, \frac{1}{s^2} \, \epsilon_W  \right) \\
\nonumber 
g^{(2)}_{HZZ}  & = &  - \, \frac{e^2}{4v} \,  \left(  \frac{s^2}{c^2}  \epsilon_{BB} \, +\, \frac{c^2}{s^2}\, \epsilon_{WW} \right) \\
\nonumber 
g^{(1)}_{HWW}  & = &   \frac{e^2}{4v}   \, \frac{1}{s^2} \, \epsilon_{W }  \\
g^{(2)}_{HWW}  & = &  - \, \frac{e^2}{2v}   \, \frac{1}{s^2} \,  \epsilon_{WW } 
\label{g1g2}
\end{eqnarray}

In addition, one obtains couplings of the Higgs H to two photons and to one photon and one $Z$-boson
\begin{eqnarray}
\nonumber
\Delta {\cal L}_{HAA}  & = & g_{HAA} \ A_{\mu \nu} A^{\mu \nu}  H  \\
\Delta {\cal L}_{HAZ}  & = & g^{(1)}_{HAZ} \ A_{\mu \nu} Z^{ \mu} \partial^\nu H   \ + \
g^{(2)}_{HAZ} \ A_{\mu \nu} Z^{ \mu \nu} H 
\label{LHVVbis}
\end{eqnarray}
where
\begin{eqnarray}
\nonumber 
g_{HAA}  & = &  - \, \frac{e^2}{4v} \,  \left(   \epsilon_{BB} \, +\, \epsilon_{WW} \right) \\
\nonumber 
g^{(1)}_{HAZ}  & = & \frac{e^2}{4v} \,  \frac{1}{s \,c} \left(  - \, \epsilon_B \, +\, \epsilon_W \right) \\
g^{(2)}_{HAZ}  & = & \frac{e^2}{2v} \,  \left(  \frac{s}{c}  \epsilon_{BB} \, - \, \frac{c}{s} \, \epsilon_{WW}  \right) 
\label{g1g2bis}
\end{eqnarray}

\subsection{The Impact of Higgs data in the effective operator basis}

The LHC measures the signal significance in each channel in terms of the {\it signal strength } $\hat{\mu}$
\bea
\hat{\mu}_i = \frac{[ \sum_j \, \epsilon_{ij} \, \sigma_{j \rightarrow H} \times {\rm Br}(H \rightarrow i)]_{observed}}{[ \sum_j \, \epsilon_{ij} \,  \sigma_{j \rightarrow H} \times {\rm Br}(H \rightarrow i)]_{SM}}\ ,
\label{mui}
\eea
where $i = 1 \cdots N_{ch}$ with $N_{ch}$  the number of channels, 
the label $j$ in the cross section, $\sigma_{j \rightarrow H}$,  is due to the fact that some final states are summed over different Higgs production processes, labelled with $j$. 
$\epsilon_{ij}$ denotes the efficiency under experimental cuts.  

\begin{figure}[h!]
\centering
\includegraphics[scale=1.5]{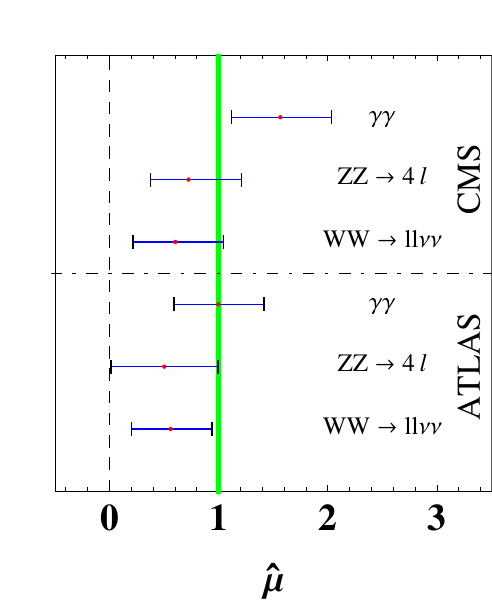}
\caption{\it Signal stregths used in this paper. The upper three correspond to CMS data, and the last three numbers are the ATLAS combination results.  }
\label{mu-fig}
\end{figure}

For this study we will use the signal strengths shown in Fig.~\ref{mu-fig}, which correspond to CMS and ATLAS combinations of 7 and 8 TeV runs, in the channels of $\gamma\gamma$, and fully leptonic $WW^{*}$ and $ZZ^{*}$. For limits from LEP, see Ref.~\cite{LEP-direct}.

\subsubsection{The diphoton channel}

The operators $\epsilon_{WW}$ and $\epsilon_{BB}$ affect the decay of $H \to \gamma\gamma$, with no relative factor. The cuts applied on the photon channel~\cite{ATLASAA, CMSAA} will not induce a difference in efficiencies when the operators are switched on because the structure of the vertex is the same, and only the overall normalization is changed. 

To do the simulation of the effective operators, we created a new model in {\tt Feynrules}~\cite{Feynrules}, adding to the SM the new operators in Eqs.~(\ref{g1g2},\ref{g1g2bis}). We then interfaced with  {\tt MadGraph}~\cite{MG5} using the UFO model format~\cite{UFO}.
We incorporated hadronization and showering effects using {\tt PYTHIA}~\cite{PYTHIA} and detector effects with 
 {\tt Delphes}~\cite{Delphes}. In our simulation, jets are always anti-k$_T$ jets of size $R=0.5$.

\begin{figure}[h!]
\centering
\includegraphics[scale=0.8]{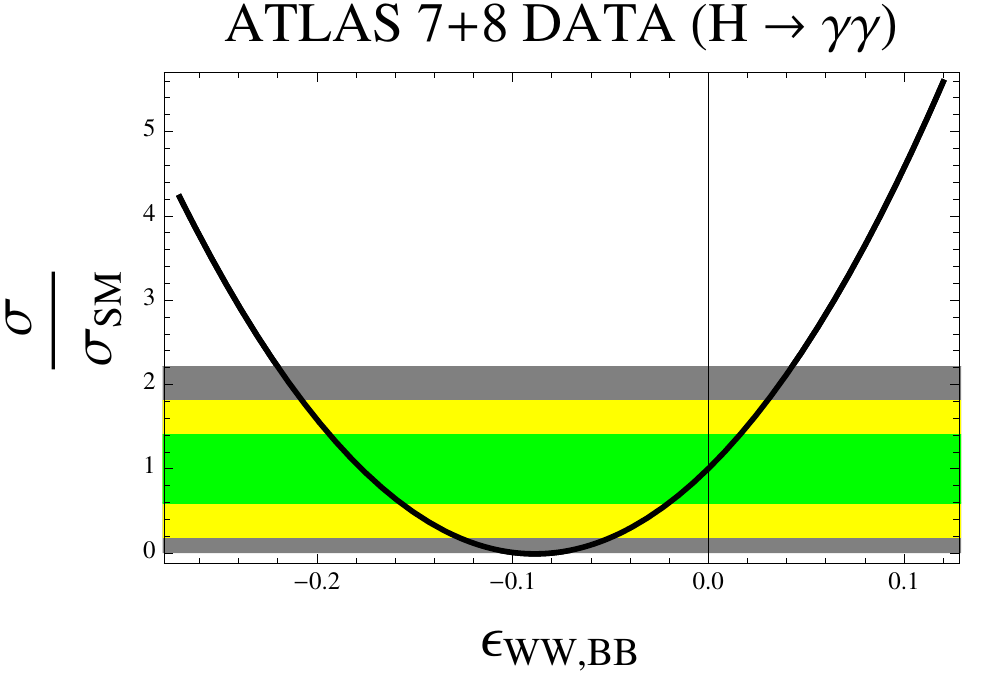}
\includegraphics[scale=0.8]{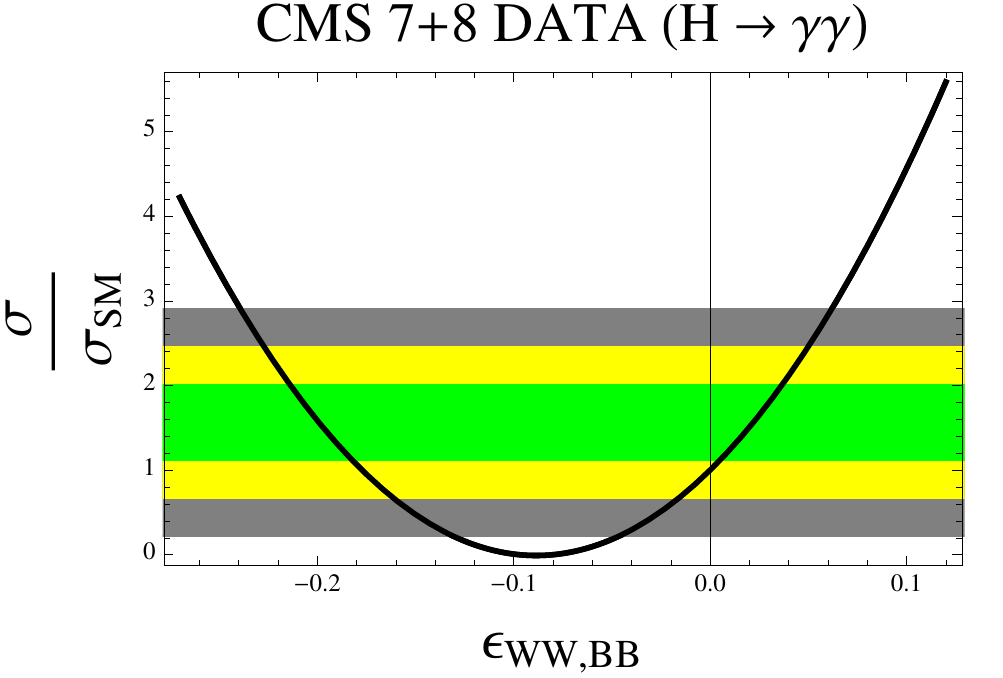}
\caption{\it . The total cross section as a function of the operators $\epsilon_{WW,BB}$ for ATLAS (left) and CMS (right) combined data.  Green, yellow and gray areas correspond to 1, 2 and 3 $\sigma$ respectively.}
\label{epsWWBB}
\end{figure}

One can extract bounds on $\epsilon_{WW}$ and $\epsilon_{BB}$ at the 95\% CL using Fig.~\ref{epsWWBB}, where the effect of the operators is shown relative to the SM. The  bounds are shown in Table~\ref{table:collbounds}, to compare with those coming from precision measurements in Sec.~\ref{LEP}, Eqs.~(\ref{LEPbounds2A} -\ref{LEPbounds2}) and (\ref{LEP2boundsA}-\ref{LEP2bounds}). The direct measurement of the Higgs to two photon surpasses the sensitivity from LEP limits on $\epsilon_{WW}$ and $\epsilon_{BB}$ by a factor ${\cal O}(10)$.

Before moving onto the other channels, let us comment about a simplification made in extracting the bounds. We are using the combined data from 7 and 8 TeV COM energies. The production cross sections are obviously slightly different, but the effect on the ratios of total cross sections, the signal strength, is negligible. Hence our simplified analysis with combined data is valid.
 
\begin{table}[t] 
\setlength{\tabcolsep}{5pt}
\center
\begin{tabular}{|c|c|c|} 
\hline \hline 
Quantity & Bound & Source
\\
\hline
 $\epsilon_{WW}$, $\epsilon_{BB}$ & [-0.21,0.03] & Diphoton-ATLAS \\
 $\epsilon_{WW}$, $\epsilon_{BB}$ & [-0.23,0.05] & Diphoton-CMS   \\
  $\epsilon_{W}$ & [-1.3,18.5] &WW-CMS and ATLAS  \\
  $\epsilon_{B}$ & $>-9.7$ & ZZ-CMS  \\
\hline \hline
\end{tabular}
\caption{\it One-parameter bounds from the $H\to \gamma\gamma$, $W W^{*}$ and $ZZ^{*}$ channels at the LHC.}
\label{table:collbounds} \vspace{-0.35cm}
\end{table}

%%%%%%%%%%%%%%%%%%%%%%%%%%
\subsubsection{The WW and ZZ channels}
%%%%%%%%%%%%%%%%%%%%%%%%%%

In the $WW$channel, the information of the angular correlation between the two leptons is used to reject background. Since our new operators in Eq.~(\ref{g1g2}) have different Lorentz structure, one could imagine a substantial difference in the angular distribution, which is indeed a way to determine the spin of the Higgs resonance. 

\begin{figure}[h!]
\centering
\includegraphics[scale=0.35]{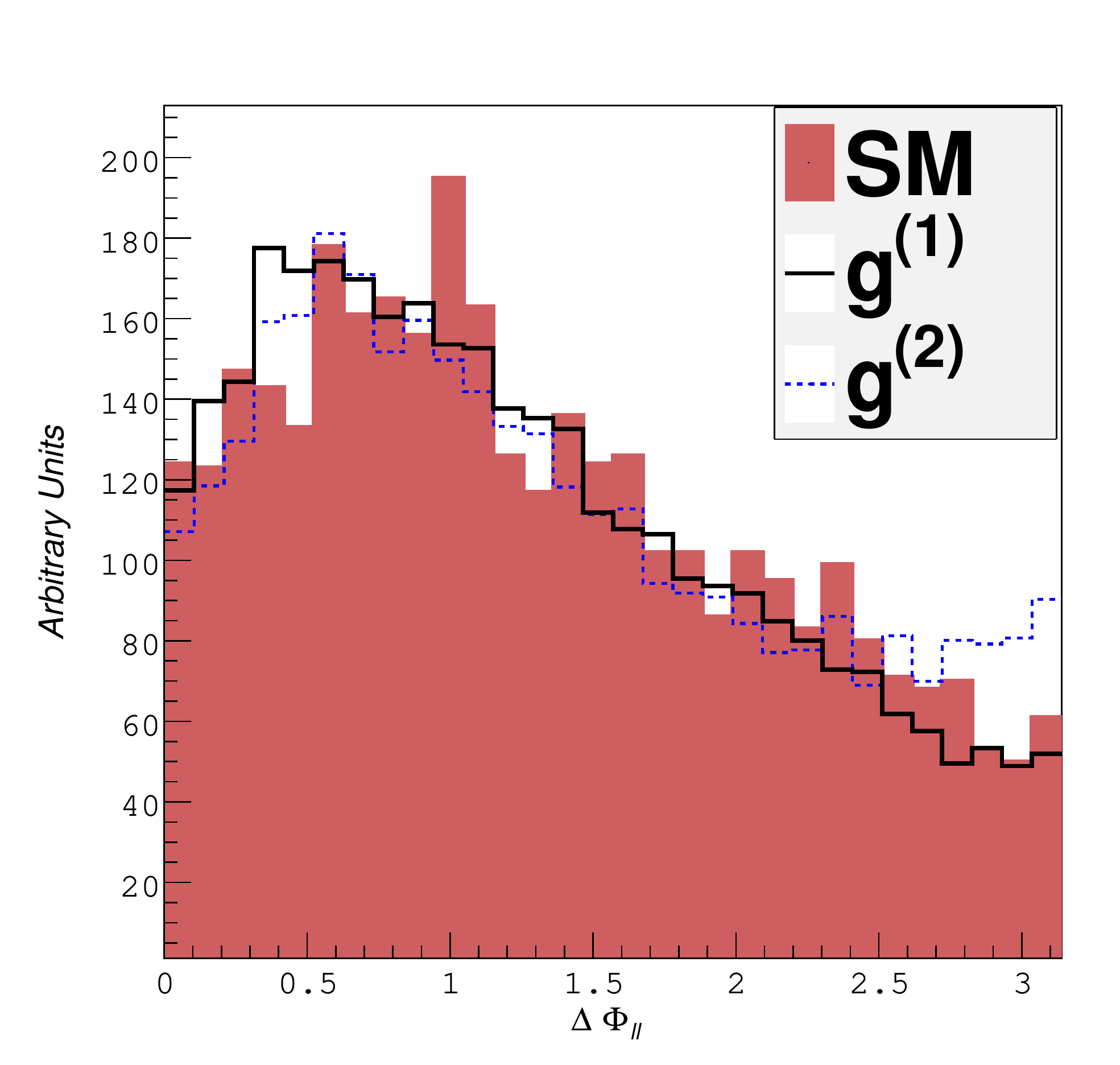}
\caption{\it The effect of the different Lorentz structures in the dilepton angular distributions. We plot the $\Delta \Phi_{\ell\ell}$ distribution for the three vertices considered here.}
\label{effW}
\end{figure}

Nevertheless we argue that, irrespective of the Lorentz structure of the vertex, the difference between the SM and the new vertex is small because it involves the spin-zero Higgs. To explain this effect, let us take the much simpler case of on-shell $WW$ production. The Higgs is a scalar, which determines the combinations of helicity in the outgoing $W$'s as $\epsilon^+_{W^-} \epsilon^-_{W^+} + \epsilon^-_{W^-} \epsilon^+_{W^+} - \epsilon^0_{W^-} \epsilon^0_{W^+}$~\cite{DaeSung}. One can then relate those polarizations to the dilepton angular distributions as shown in Ref.~\cite{DaeSung}, to find that the distribution in terms of the azimuthal angle difference ($\Delta \Phi_{\ell\ell}$) is  a decreasing funtion.  As we move into the real situation, where at least one of the $W$'s is off-shell, this behavior  qualitatively persists. This can be seen in an explicit simulation of the effect of the different vertices, as shown in Fig.~\ref{effW}. We plotted the angular distribution of the dilepton system when vertices of the SM, $g^{(1)}$ and $g^{(2)}$ types are switched on. The distributions are very similar because the fact that the leptons tend to be produced in parallel is a consequence of the spin of the Higgs. Note, though, that one could try to extract the Lorentz structure of the vertex in a linear collider~\cite{TaoHan, Takubo}, and in the vector boson fusion channel~\cite{Hankele}, or possibly with more data~\cite{Plehn,Stolarski}. 

Although the differences are small,  we would like to quantify them by implementing the ATLAS~\cite{ATLASWW} and CMS~\cite{CMSWW} searches for the Higgs to two leptons.
Our implementation of the ATLAS analysis starts with a  selection of events with two opposite-sign, opposite-flavour leptons with $p_T^{\ell_1,\ell_2}>$ 25,15 GeV in the central region, and invariant mass 50 GeV $> m_{\ell\ell} >$ 10 GeV. Quality and isolation criteria are applied at the level of Delphes simulation. We will focus on the  zero- and one- and two-jet analysis. The jets are asked to have $p_T>$ 25, 30 GeV in the central, forward region.
In the zero-jet region, the final cuts applied are 
\bea
E^{miss}_{T,rel}> 25 \textrm{ GeV, }  p_{T}^{\ell\ell} > 30 \textrm{ GeV, and }  |\Delta \Phi_{\ell\ell}| < 1.8 
\eea
whereas in the one-jet  case, there is an extra cut
\bea
\vec{p}_{T}^{\ell\ell}+ \vec{p}_{T}^{j}+\vec{E}_{T}^{miss} < 30 \textrm{ GeV}
\eea
besides a b-tag veto.  Finally, a cut on $m_{T}$ between 93.75 and 125 GeV is applied.

We also simulate the corresponding CMS search . CMS cuts are very similar to ATLAS, but now $p_T^{\ell\ell}> $ 45 GeV, $m_{\ell\ell }\in [12,45]$ GeV, $\Delta \Phi_{\ell\ell}< 1.6$ and $m_{T}\in [80,125]$ GeV. Let us note that the $\Delta \Phi_{\ell\ell}$ cut is correlated to the other two cuts.

\begin{figure}[h!]
\centering
\includegraphics[scale=0.8]{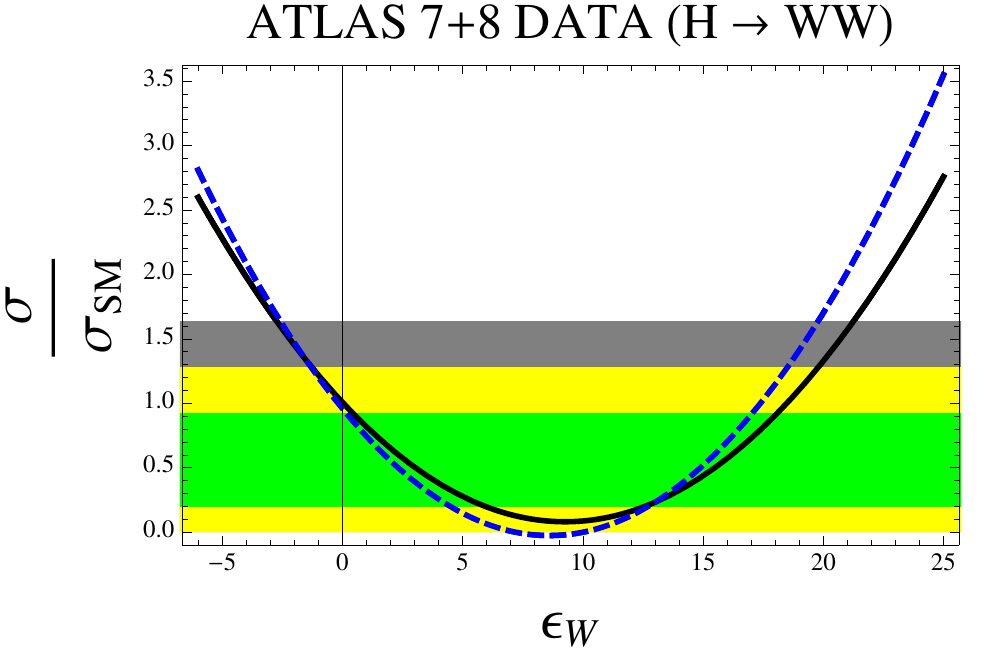}
\includegraphics[scale=0.8]{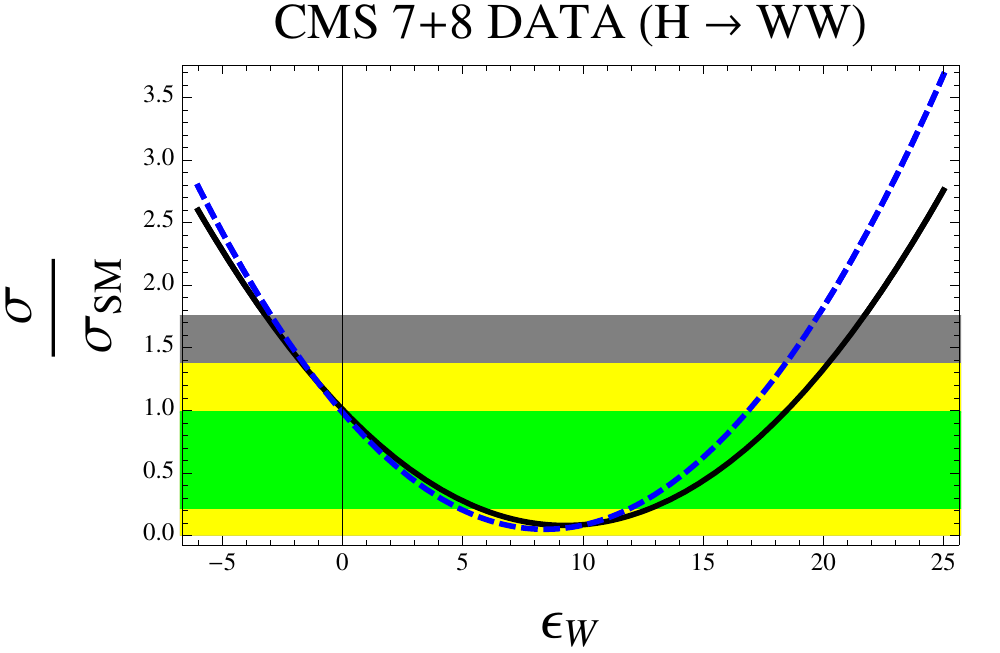}
\caption{\it . The total cross section as a function of the operator $\epsilon_{W}$ for ATLAS (left) and CMS (right) combined data.  Green, yellow and gray areas correspond to 1, 2 and 3 $\sigma$ respectively. The blue-dashed line corresponds to having the efficiencies effect into account.}
\label{epsW}
\end{figure}

We present our results in Fig.~\ref{epsW} for the case of the operators $\epsilon_{W,B}$, which generate couplings of the type $g^{(1)}$. We do not use the $WW$ channel to constrain the $\epsilon_{WW,BB}$ operators, as the $\gamma\gamma$ limits are much better. The black line is the value of $\hat{\mu}$ without cuts. The dashed-blue line corresponds to the same quantity with cuts taken into account. The slight difference among lines reflects the little distinction between the SM and the $g^{(1)}$ type of couplings in the cuts applied. 

Finally, we looked into the Higgs to four leptons, via $Z Z^*$. ATLAS~\cite{ATLASZZ} and CMS~\cite{CMSZZ} use quite different techniques for the time being. ATLAS is a cut based analysis, which essentially asks for $m_{2\ell}$ and $m_{4\ell}$ in the range of the $Z$ and $H$ masses. CMS uses a rather sophisticated multivariate analysis called MELA, which is based in four angular observables and one invariant mass. As we discussed for the $WW^*$ case, we do not expect any sizable effect on efficiencies due to the new couplings, less so in the case of ATLAS $ZZ^*$ analysis. In the global fit presented later on, we will use information from all these channels, but the $\gamma\gamma$ will be the observable from the LHC leading the constraints, with the $WW$ and $ZZ$ channels, limited by statistics, playing a less important role.

%%%%%%%%%%%%%%%%%%%%%%%
\section{Combined Constraints on Anomalous Couplings}\label{combined}
%%%%%%%%%%%%%%%%%%%%%%%

Now we would like to repeat the exercise we did at the end of Sec.~\ref{LEP},
namely, to find the constraints in the planes $(\epsilon_{WW}, \epsilon_{BB})$ and $(\epsilon_{W}, \epsilon_{B})$.
Of course, it is expected that adding the recent LHC data will improve these constraints, but the degree of amelioration
is something we would like to evaluate in this Section. 

\begin{figure}[h!]
\centering
\includegraphics[scale=1.]{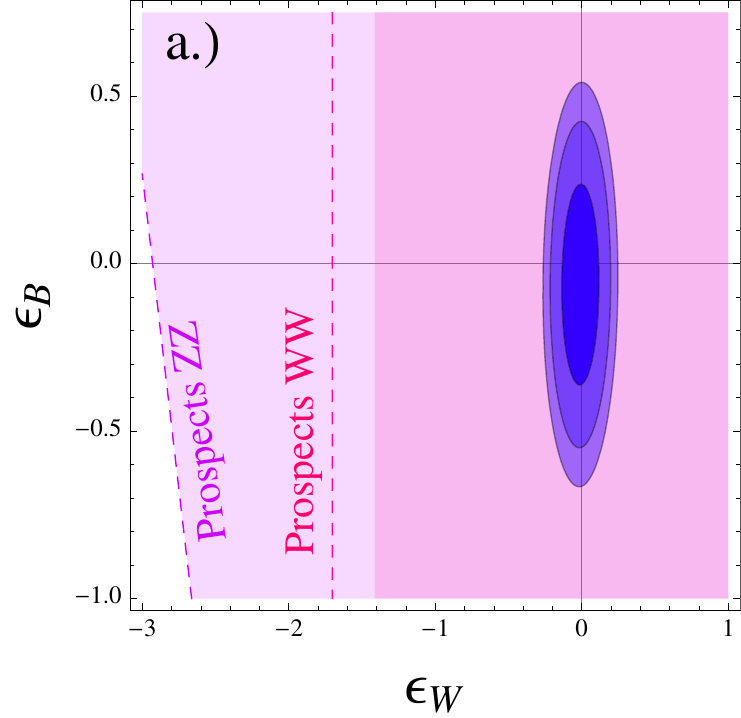}
\hspace{.5cm}
\includegraphics[scale=1.]{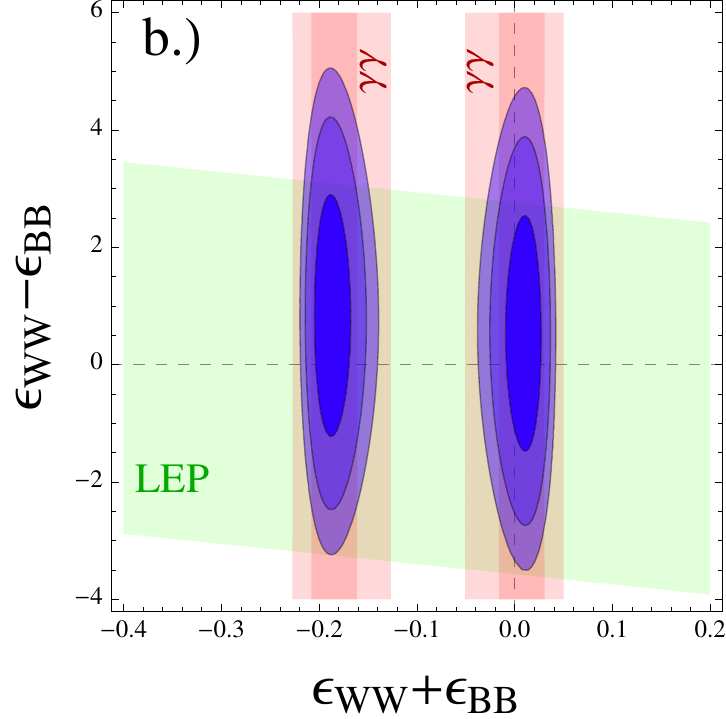}
\caption{\it . Constraints from electroweak precision data and LHC data on the coefficients $(\epsilon_{W}, \epsilon_{B})$ and
$(\epsilon_{WW} + \epsilon_{BB}, \epsilon_{WW} - \epsilon_{BB})$. 
In blue, the 99, 95 and 68\%CL combined constraints. We show the individual constraints coming from LEP and from LHC
(regions corresponding to $\Delta \chi^2$ = 3.84)
 and the prospects (see text).}
\label{epstot}
\end{figure}

In Fig.~\ref{epstot}a we see the results on $(\epsilon_{W}, \epsilon_{B})$. Comparing with Fig.~\ref{STfigure}a we
notice there is no substantial improvement with the LHC so far, to the extent that the ellipse from LEP and from the combined LEP+LHC are basically the same. In Sec.~\ref{prospects} we will explain why we do not expect  much improvement from the full 2012 LHC dataset for those parameters.

In Fig.~\ref{epstot}b  we see the results on  $(\epsilon_{WW}, \epsilon_{BB})$;
in this case the improvement is dramatic since we go from constraints in form of stripes in Fig.~\ref{STfigure}b to contour
ellipses in Fig.~\ref{epstot}b. Since the constraints on $(\epsilon_{WW}$ and  $\epsilon_{BB})$ are dominated by the $\gamma\gamma$ channel,
the constraints are better presented in terms of the orthogonal combinations $\epsilon_{WW} + \epsilon_{BB}$ 
and $\epsilon_{WW} -\epsilon_{BB}$. Also, due to the fact that the present $\gamma\gamma$ data exceed the theoretical SM
prediction by more than one-sigma there are actually two constrained regions in the Fig\ref{epstot}b.

In Fig.~\ref{epstot} we also show the constraints coming from LEP alone and the ones coming from the different LHC channels.
We hope this clarifies even more the role of the separate experimental constraints.
In Fig.~\ref{epstot}b we show how the stripe coming from LEP (basically the one of Fig.~\ref{STfigure}b) nicely complements
the stripe coming from the $\gamma\gamma$ LHC channel. The narrower, darker, horizontal stripe is the CMS data,
the wider, lighter, stripe is the ATLAS data.
We do not show the stripes corresponding to $WW$ and $ZZ$ LHC
channels because they are much more loose, and actually the stripe borders are outside the region of parameters we display
in the Fig.~\ref{epstot}b. In Fig.~\ref{epstot}a we show the role of the $WW$ LHC channel (darker region) and the $ZZ$ LHC channel 
(lighter region).

%%%%%%%%%%%%%%%%%%%%%%%
\section{Limits on the Coupling of the Higgs to Photon-Z}\label{gHAZsec}
%%%%%%%%%%%%%%%%%%%%%%%

Higher order operators can induce a coupling of the Higgs to a Z and a photon, as shown in Eqs.~\ref{g1g2bis}. The decay rate is then given by
\bea
\Gamma (H \to \gamma Z) = \frac{m_H^3}{16 \pi} \, \left( 1- \frac{m_Z^2}{m_H^2}\right)^3 \, |g_{HAZ}^{(1)}+2 g_{HAZ}^{(2)}+\kappa_{SM}|^2
\eea
The SM contribution is  $\kappa_{SM} \simeq - 4.1 \times 10^{-5}$ GeV$^{-1}$, where we have included the $W$ and top loops. For a discussion on the expected LHC sensitivity to direct measurements, see Ref.~\cite{Ian-photonZ}.

\begin{figure}[h!]
\centering
\includegraphics[scale=.75]{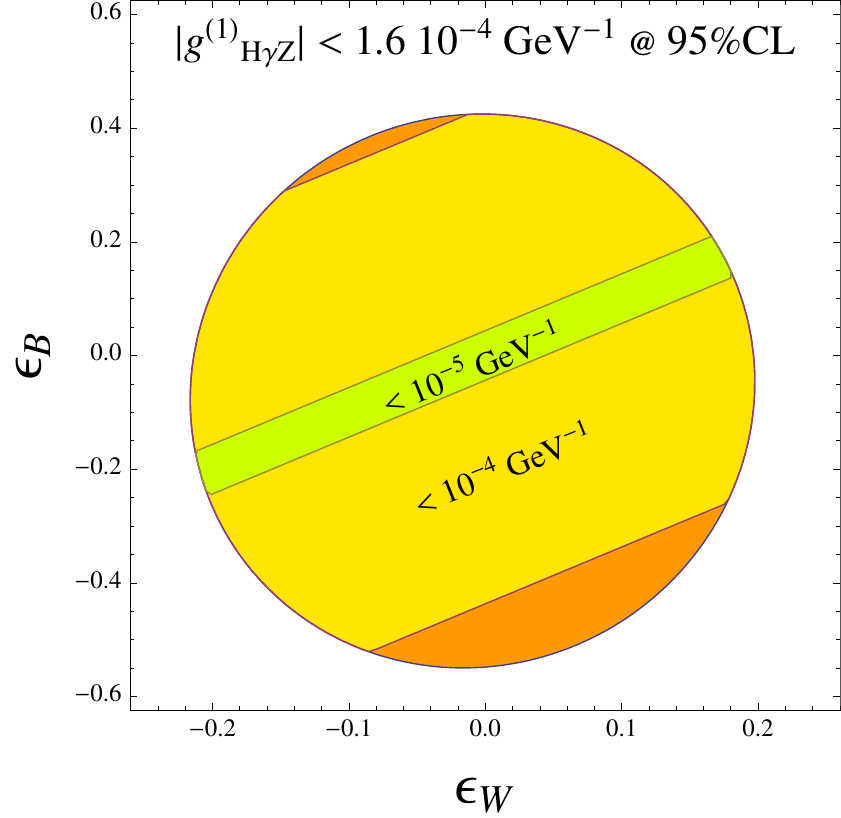}
\includegraphics[scale=.72]{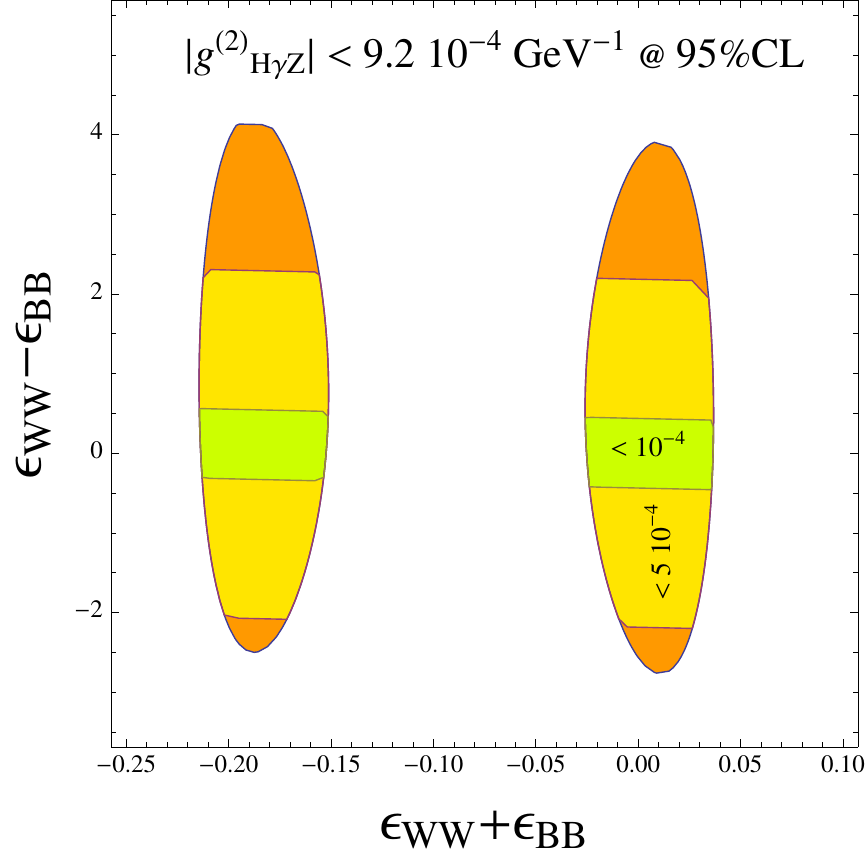}
\caption{\it . Limits on the coupling of the Higgs to $\gamma Z$ in the ($\epsilon_W$, $\epsilon_B$) (left) and  ($\epsilon_{WW}$, $\epsilon_{BB}$) (right) parameter space. The current limit on the generated couplings $g^{(1,2)}$ is shown in the plot, as well as slices of tighter upper limits. }
\label{g1HAZ}
\end{figure}

In Fig.~\ref{g1HAZ} we show lines of fixed bounds for $g_{HAZ}^{(1)}$ and $g_{HAZ}^{(2)}$ in the  the ($\epsilon_{WW}$,$\epsilon_{BB}$) and ($\epsilon_{W}$,$\epsilon_{B}$) parameter space, respectively. One can infer then limits in the 95\% CL which read 
\bea
|g_{HAZ}^{(1)}| & < & 1.6 \times 10^{-4} \textrm{ GeV}^{-1} \\
|g_{HAZ}^{(2)}| & < & 9.2 \times 10^{-4} \textrm{ GeV}^{-1}
\eea
or, equivalently
\bea
\Gamma  (H \to \gamma Z)  < 1.0 \times 10^{-4} \, ( 1.3 \times 10^{-2} ) \,  \textrm { GeV} 
\eea
depending on which operator is switched on, $g^{(1)}$ ($g^{(2)}$)~\footnote{Note that our limits on $g^{(2)}$ are more conservative than those in Ref.~\cite{Hankele}.}.  Note that the SM contribution is very small, $\Gamma (H \to Z \gamma)_{SM}\simeq 6\times 10^{-6}$ GeV. In Fig.~\ref{g1HAZ} we also show contours  of $g_{HAZ}^{(1)}< 10^{-4}$, $10^{-5}$ GeV$^{-1}$ and $g_{HAZ}^{(2)}< 5 \times 10^{-4}$, $10^{-4}$ GeV$^{-1}$. 

\begin{figure}[h!]
\centering
\includegraphics[scale=.25]{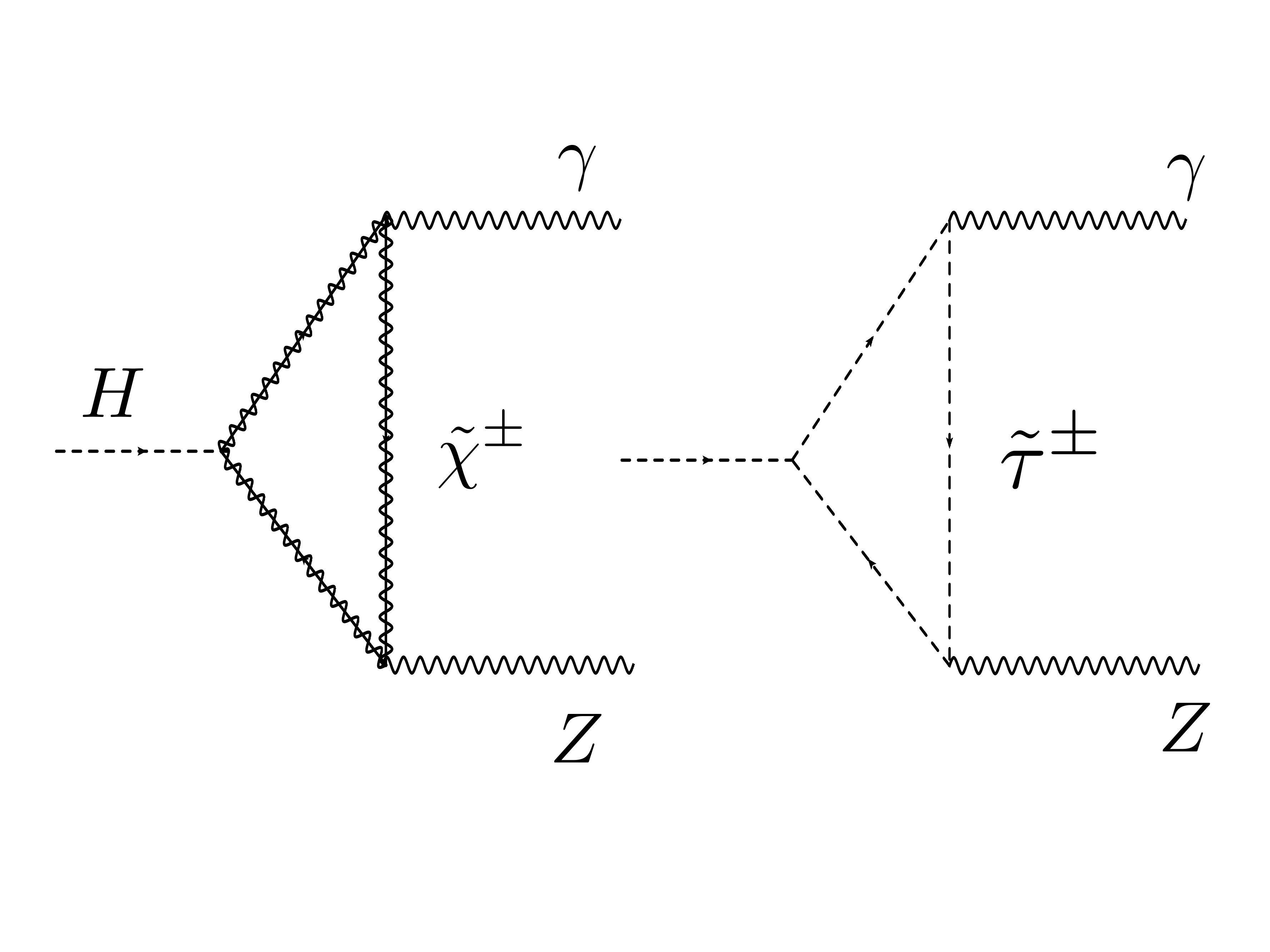}
\caption{\it . Chargino/Stau contribution to loop diagrams leading to the anomalous coupling $A_{\mu\nu} Z^{\mu\nu} H$. }
\label{feyn-char}
\end{figure}

One could interpret the bounds on the decay in terms of new physics generated by, for example, Supersymmetry. A coupling of the kind $g^{(2)} H A_{\mu\nu} Z^{\mu\nu}$
could be generated by a loop of charged Higgses, charginos and staus as shown in the diagram of Fig.~\ref{feyn-char}~\cite{revisited}. We consider the interpretation in terms of exclusively  electroweak states, neglecting the effect of stops,  which should be heavy enough as to not influence this vertex, and neither the gluon coupling to Higgses.
We will also work in the decoupling limit of the two-Higgs doublet model, as the observed Higgs and the bounds on the pseudo-scalar Higgs are consistent with this assumption. In this case, the charged Higgs contribution is two orders of magnitude smaller than the $W$ contribution ($g^{(2)}_{H^{\pm}}=\alpha/(24 \sqrt{2} \pi  s c v )\simeq 7 \times 10^{-7}$ Gev$^{-1}$), hence this analysis provides no information on the charged Higgs. 

In the limit of heavy chargino, its contribution would be given by 
\bea
g_{HAZ}^{(2)} (\tilde{\chi}^{\pm}) = -\frac{\alpha}{3 \sqrt{2} s v} \frac{m_Z}{m_{\tilde{\chi}^{\pm}}} \, g_{Z \tilde{\chi}^{\pm} \tilde{\chi}^{\mp}} \, g_{H \tilde{\chi}^{\pm} \tilde{\chi}^{\mp}}
\eea
where $g_{(Z,H) \tilde{\chi}^{\pm} \tilde{\chi}^{\mp}}$ are the couplings of the $Z$ and $H$ to the charginos, and they are bounded by $\lesssim1$. This implies that the chargino contribution is about one order of magnitude smaller than the SM one for $m_{\tilde{\chi}^{\pm}} \gtrsim v$.

The situation for staus is more promising, provided the staus have a large LR mixing. In this limit the induced coupling can be estimated as
\bea
g_{HAZ}^{(2)} (\tilde{\tau}) \simeq -\frac{\alpha}{6 \sqrt{2} s c v} \left(\frac{m_{LR}}{m_{\tilde{\tau}_{\ell}}}\right)^2 \,
\eea
where $\tilde{\tau}_{\ell}$ is the lightest stau, and $m_{LR}$ is the off-diagonal mass term, $m_{\tau} (A_{\tilde{\tau}}+\mu/\tan_{\beta})$, and we have taken the limit in the loop integrals of heavy stau. 
The limit on the anomalous coupling can then be translated in
\bea
m_{LR} \lesssim 30 \, m_{\tilde{\tau}_{\ell}} \ .
\eea
 
By going through this exercise, we see that the current dataset it not sensitive to charginos and charged Higgs via the indirect probes we are discussing here~\cite{haa}. In the case of the stau, the sensitivity depends largely on the amount of mixing between the two staus. But note that a large $m_{LR}/m_{\tilde{\tau}_{\ell}}$ means that there would be a large gap between the two stau physical states, possibly getting into the dangerous region of charge breaking minima if $m_{\tilde{\tau}_{\ell}}^2<$0. 
 
%%%%%%%%%%%%%%%%%%%%%%%
\section{Constraining New Physics: A Toy Example}\label{toy}
%%%%%%%%%%%%%%%%%%%%%%%
As a final illustration of the effect of the bounds on effective operators on UV models, we use the example of a radion in warped extra-dimensions. We follow closely the discussion in Ref.~\cite{Csaba-Jay}. 

The coupling of the radion $R$ to massless gauge fields is loop induced and given by
\bea
{\cal L } \supset\frac{R}{L \, \Lambda}  \, \left(\frac{1}{g^2} \widehat  W^{\mu\nu} \widehat W_{\mu\nu} +   \frac{1}{g'^2} \widehat  B^{\mu\nu} \widehat B_{\mu\nu} \right) 
\label{radionFF}
\eea
where we are neglecting the localized kinetic terms and trace anomalies. Here $L=\log(M_P/TeV)\simeq 30$ in the usual Randall-Sundrum (RS) model, but can be smaller in the Little Randall-Sundrum (LRS)~\cite{little-RS}. For example, with a cutoff of order 100 TeV, one would expect $L\simeq 5$. Note that after EWSB, this coupling receives an extra contribution, suppressed by order the effective volume of the extra-dimension, i.e. ${\cal O} (30)$. 

Assuming there is no Higgs-curvature mixing, and that the Higgs is localized on the IR brane, the coupling of the radion to the Higgs would be given by
\bea
{\cal L } \supset \frac{2 R}{\Lambda} \, m_H^2  \Phi^{\dagger} \Phi
\eea

Integrating out the heavy radion, leads to the effective operator
\bea
\epsilon_{BB} \simeq - \left(2 \frac{m_H v}{\Lambda m_R} \right)^2 \, \frac{1}{L g'^2} =   \frac{g^2}{g'^2} \epsilon_{WW}
\eea

\begin{figure}[h!]
\centering
\includegraphics[scale=1.4]{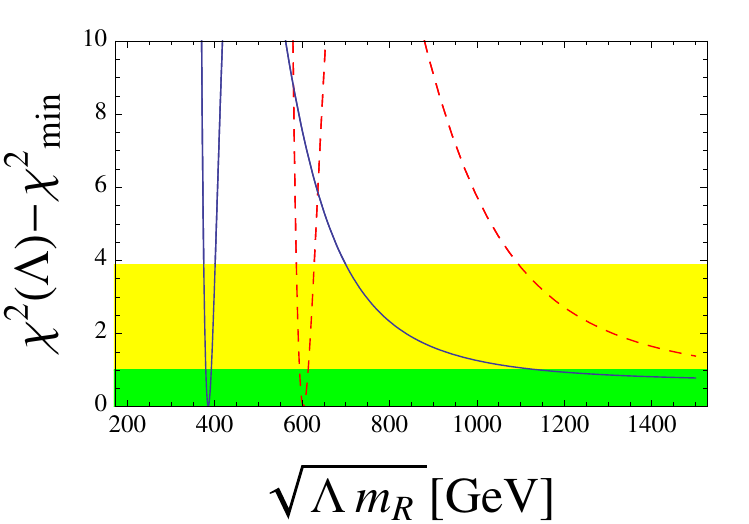}
\caption{\it The limit on the mass scale $\sqrt{\Lambda m_R}$ for the bulk RS with $L$=30 (blue-solid line) and LRS ($L=5$) (red-dashed line).}
\label{radion}
\end{figure}

The typical energies at which we are probing those anomalous couplings are $Q^2 \simeq m_H^2$, hence in our expansion of the radion dynamics, we are neglecting terms of order $m_H^2/m_R^2$.  In Fig.~\ref{radion}, we present the limit in terms of $\sqrt{\Lambda m_R}$, for the RS (LRS) model. The current limit on $\sqrt{ \Lambda m_R}$ is about 700 (1100) GeV, barring possible tuning into the tiny region at low mass.

Note that the radion could also couple to the gluon, and modify the Higgs production mechanism. However, for $SU(3)$, the trace anomaly becomes an important contribution, and  the coupling is modified by $\propto (1- \alpha_s b L/(2 \pi))$. Here $b$ is the total beta function of order {\cal{O}(1 - 10)}, depending on the localization of colored fields in the extra-dimension. In this case, a partial cancellation in the radion coupling to gluons is conceivable, and could reduce its effect on the gluon fusion process.

%%%%%%%%%%%%%%%%%%%%%%%
\section{Prospects for the 2012 run}\label{prospects}
%%%%%%%%%%%%%%%%%%%%%%%

In this section we look ahead into the end of the year's run, and how the full dataset would affect the results presented in this paper. Lacking a crystal ball, we must introduce some theoretical bias. Our choice is to assume that the central values on the signal strengths $\hat{\mu}$ will move towards the SM expected values, i.e. we shall set $\hat{\mu}=1$. Moreover, we will estimate that the error bars would go down by a factor of two, assuming a total of 30$^{-1}$ fb of data. This is most probably a very optimistic estimate, as some of the signal strengths are not lying around the SM value, and in migrating towards it, the errors are not expected to scale so quickly. Nevertheless, this exercise allows us to illustrate the impact of more precise data in the current analysis.

The effect on the $(\epsilon_W, \epsilon_B)$ is minimal, as we display in Fig.~\ref{epstot}a. As the central values on $WW^*$ and $ZZ^*$ channels move towards the SM, the reduction of the error bars is barely affecting the global fit. 

\begin{figure}[h!]
\centering
\includegraphics[scale=1.]{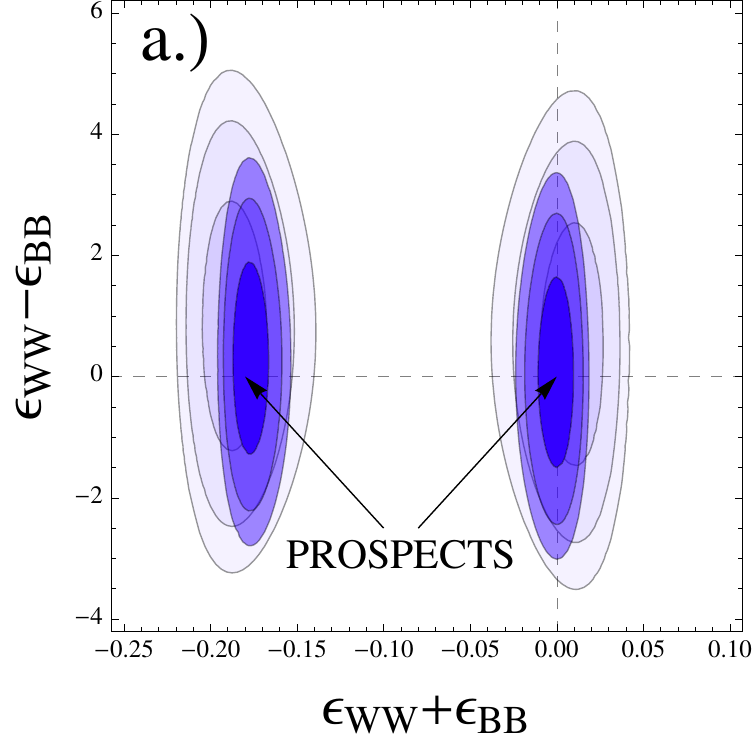}
\hspace{0.5cm}
\includegraphics[scale=.95]{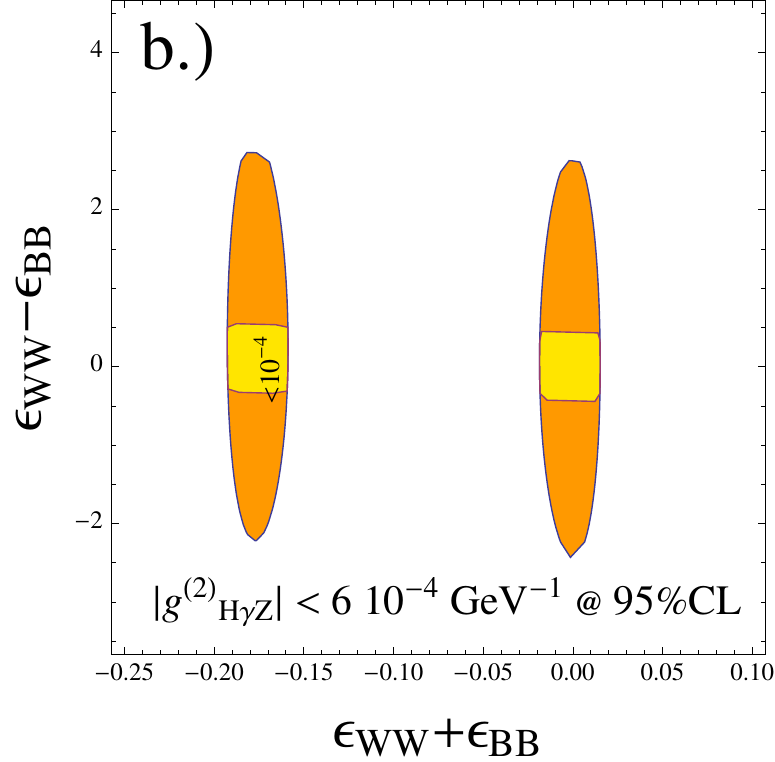}
\caption{\it }
\label{prospects-fig}
\end{figure}

The situation for the  $(\epsilon_{WW}, \epsilon_{BB})$ is more encouraging. In Fig.~\ref{prospects-fig}a we show the improvement in the fit, and in Fig.~\ref{prospects-fig}b, the improvement on the limits on $g_{HAZ}$, both a factor ${\cal O}$(1.5-2).

%%%%%%%%%%%%%%%%%%%%%%%
\section{Conclusions}\label{conclusions}
%%%%%%%%%%%%%%%%%%%%%%%

In this paper we have taken the approach that the Higgs candidate is an elementary scalar  and that the leading effects of New Physics appear at the level of dimension-six effective operators. We focused on four operators which affect the couplings to electroweak gauge bosons, $W$, $Z$ and $\gamma$, to constrain deviations from the SM behavior, which we named $\epsilon_{W}$, $\epsilon_{B}$, $\epsilon_{WW}$ and $\epsilon_{BB}$.

We have constrained those parameters one by one, and also by pairs ( $\epsilon_{W,B}$ and  $\epsilon_{WW,BB}$), as  in standard scenarios of UV completions those tend to come together.  

We started by looking at constraints from LEP1 (and low energy electroweak data) and LEP2, which are especially restrictive for the $\epsilon_{W,B}$ operators. After LHC data is taken into account, limits on $\epsilon_{W,B}$ do not improve significantly, and we expect no sizeable improvement with the full 2012 dataset.

On the other hand, the operators  $\epsilon_{WW,BB}$, poorly constrained by LEP data, contribute to the Higgs to two photon coupling. The sensitivity with the current LHC data is better than LEP by a factor ${\cal O}(10)$. With more LHC data coming, we estimate those limits will improve by a factor of around two. 

We studied the impact of non-standard Lorentz structures in the coupling of the Higgs to $WW$. As the $WW^{*}$ experimental analysis makes use of angular correlations between the two leptons in the $W$ decays, one could expect a modification of the efficiency to the cut on dilepton azimuthal angle. We found the effect is negligible, something one can qualitatively understand by realizing that the Higgs would predominantly produce parallel leptons.

We then performed a combined fit to LEP and LHC data, with no significant changes respect to what we have already obtained using the individual channels. Namely,  limits on  $\epsilon_{W,B}$ driven by LEP, whereas  $\epsilon_{WW,BB}$ is mostly determined by the LHC gamma-gamma signal. 

One particularly interesting anomalous coupling is the rare decay $H\to Z \gamma$. Both sets of operators can induce this coupling. We show that the limits on the decay width are at least an order of magnitude larger than the SM prediction. We then interpreted the limits in Supersymmetry with light electroweak states, charged Higgses, charginos and staus, to find that this dataset is only sensitive to staus, possibly with large mixing.

Besides Supersymmetry,  extra-dimensional scenarions are a possible source of these operators. We have discussed the effect of the exchange of a massive radion, and set limits in terms of its mass and scale of interaction of the order of the TeV.

%%%%%%%%%%%%%%%%%%%%%%%%
\appendix*

\subsection*{Acknowledgments}

We thank Abdelhak Djouadi, John Ellis, Jens Erler, Belen Gavela, Gian Giudice, Concha Gonzalez-Garcia, Christophe Grojean, Ramon Miquel, Manuel Perez-Victoria, Alex Pomarol, and Mike Trott for enlightening discussions.
EM would like to thank CERN for hospitality while part of this work was done, 
and the partial support from the projects FPA2011-25948, 2009SGR894.

%%%%%%%%%%%%%%%%%%%%%%%%%%%%%%%%%%%%%%%%%%


\begin{thebibliography}{99}
%%%%%%%%%%%%%%%%%%%%%%%%%%%%%%%%%%%%%%%%%%%%%%%%%%%%%%%%%%%
%%%%%%%%%%%%%


 \bibitem{:2012gk} 
  G.~Aad {\it et al.}  [ATLAS Collaboration],
  %``Observation of a new particle in the search for the Standard Model Higgs boson with the ATLAS detector at the LHC,''
  Phys.\ Lett.\ B {\bf 716}, 1 (2012)
  [arXiv:1207.7214 [hep-ex]].
  S.~Chatrchyan {\it et al.}  [CMS Collaboration],
  %``Observation of a new boson at a mass of 125 GeV with the CMS experiment at the LHC,''
  Phys.\ Lett.\ B {\bf 716}, 30 (2012)
  [arXiv:1207.7235 [hep-ex]]. 
  
  \bibitem{fits}
    J.~R.~Espinosa, C.~Grojean, M.~Muhlleitner and M.~Trott,
  %``Fingerprinting Higgs Suspects at the LHC,''
  JHEP {\bf 1205}, 097 (2012)
  [hep-ph/1202.3697].
  J.~Ellis and T.~You,
  %``Global Analysis of Experimental Constraints on a Possible Higgs-Like Particle with Mass ~ 125 GeV,''
  JHEP {\bf 1206}, 140 (2012)
  [hep-ph/1204.0464]; [hep-ph/1207.1693].
  J.~Ellis and T.~You,
  arXiv:1207.1693 [hep-ph].
  %%CITATION = ARXIV:1207.1693;%% 
  %%CITATION = ARXIV:1204.0464;%%
  %%CITATION = ARXIV:1202.3697;%%
  T.~Corbett, O.~J.~P.~Eboli, J.~Gonzalez-Fraile and M.~C.~Gonzalez-Garcia,
  %``Constraining anomalous Higgs interactions,''
   J.~R.~Espinosa, C.~Grojean, M.~Muhlleitner and M.~Trott,
  %``Probing for Invisible Higgs Decays with Global Fits,''
  [hep-ph/1205.6790].
  %%CITATION = ARXIV:1205.6790;%%
  arXiv:1207.1344 [hep-ph]. J.~R.~Espinosa, C.~Grojean, M.~Muhlleitner and M.~Trott,
  %``First Glimpses at Higgs' face,''
  [hep-ph/1207.1717].
  %%CITATION = ARXIV:1207.1717;%% 
   A.~Azatov, R.~Contino and J.~Galloway,
  %``Model-Independent Bounds on a Light Higgs,''
  JHEP {\bf 1204}, 127 (2012)
  [arXiv:1202.3415 [hep-ph]].
  %%CITATION = ARXIV:1202.3415;%%
   M.~Montull and F.~Riva,
  %``Higgs discovery: the beginning or the end of natural EWSB?,''
  arXiv:1207.1716 [hep-ph].
  %%CITATION = ARXIV:1207.1716;%%
   M.~Klute {\it et al.}
  %``Measuring Higgs Couplings from LHC Data,''
  [hep-ph/1205.2699].
  %%CITATION = ARXIV:1205.2699;%%
 A.~Azatov, R.~Contino and J.~Galloway,
  %``Contextualizing the Higgs at the LHC,''
  [hep-ph/1206.3171].
  %%CITATION = ARXIV:1206.3171;%%


\bibitem{HQN}
  Y.~Gao, A.~V.~Gritsan, Z.~Guo, K.~Melnikov, M.~Schulze and N.~V.~Tran,
  %``Spin determination of single-produced resonances at hadron colliders,''
  Phys.\ Rev.\ D {\bf 81} (2010) 075022
  [arXiv:1001.3396 [hep-ph]].
  %%CITATION = ARXIV:1001.3396;%%
  S.~Y.~Choi, D.~J.~Miller, 2, M.~M.~Muhlleitner and P.~M.~Zerwas,
  %``Identifying the Higgs spin and parity in decays to Z pairs,''
  Phys.\ Lett.\ B {\bf 553} (2003) 61
  [hep-ph/0210077].
  %%CITATION = HEP-PH/0210077;%%
J.~Ellis and D.~S.~Hwang,
  %``Does the `Higgs' have Spin Zero?,''
  arXiv:1202.6660 [hep-ph].
  %%CITATION = ARXIV:1202.6660;%%
   A.~De Rujula, J.~Lykken, M.~Pierini, C.~Rogan and M.~Spiropulu,
  %``Higgs look-alikes at the LHC,''
  Phys.\ Rev.\ D {\bf 82} (2010) 013003
  [arXiv:1001.5300 [hep-ph]].
K.~Odagiri,
  %``On azimuthal spin correlations in Higgs plus jet events at LHC,''
  JHEP {\bf 0303} (2003) 009
  [arXiv:hep-ph/0212215].
  C.~P.~Buszello, I.~Fleck, P.~Marquard and J.~J.~van der Bij,
  %``Prospective Analysis Of Spin- And Cp-Sensitive Variables In H $\to$ Z Z
  %$\to$ L(1)+ L(1)- L(2)+ L(2)- At The Lhc,''
  Eur.\ Phys.\ J.\  C {\bf 32} (2004) 209
  [arXiv:hep-ph/0212396].
  A.~Bredenstein, A.~Denner, S.~Dittmaier and M.~M.~Weber,
  %``Precise predictions for the Higgs-boson decay H --> W W / Z Z --> 4
  %leptons,''
  Phys.\ Rev.\  D {\bf 74} (2006) 013004
  [arXiv:hep-ph/0604011].
   P.~S.~Bhupal Dev, A.~Djouadi, R.~M.~Godbole, M.~M.~Muhlleitner and S.~D.~Rindani,
  %``Determining the CP properties of the Higgs boson,''
  Phys.\ Rev.\ Lett.\  {\bf 100} (2008) 051801
  [arXiv:0707.2878 [hep-ph]].
   U.~De Sanctis, M.~Fabbrichesi and A.~Tonero,
  %``Telling the spin of the 'Higgs boson' at the LHC,''
  Phys.\ Rev.\  D {\bf 84} (2011) 015013
  [arXiv:1103.1973 [hep-ph].
   J.~Ellis, V.~Sanz and T.~You,
  %``Prima Facie Evidence against Spin-Two Higgs Impostors,''
  arXiv:1211.3068 [hep-ph].
  %%CITATION = ARXIV:1211.3068;%%
  %10 citations counted in INSPIRE as of 28 Jun 2013
  R.~Boughezal, T.~J.~LeCompte and F.~Petriello,
  %``Single-variable asymmetries for measuring the `Higgs' boson spin and CP properties,''
  arXiv:1208.4311 [hep-ph].
  %%CITATION = ARXIV:1208.4311;%%
  D.~Stolarski and R.~Vega-Morales,
  %``Directly Measuring the Tensor Structure of the Scalar Coupling to Gauge Bosons,''
  arXiv:1208.4840 [hep-ph].
  J.~Ellis, D.~S.~Hwang, V.~Sanz and T.~You,
  %``A Fast Track towards the `Higgs' Spin and Parity,''
  arXiv:1208.6002 [hep-ph].
   A.~Djouadi, R.~M.~Godbole, B.~Mellado and K.~Mohan,
  %``Probing the spin-parity of the Higgs boson via jet kinematics in vector boson fusion,''
  Phys.\ Lett.\ B {\bf 723} (2013) 307
  [arXiv:1301.4965 [hep-ph]].
  %%CITATION = ARXIV:1301.4965;%%
  %10 citations counted in INSPIRE as of 28 Jun 2013
   R.~Godbole, D.~J.~Miller, K.~Mohan and C.~D.~White,
  %``Boosting Higgs CP properties via VH Production at the Large Hadron Collider,''
  arXiv:1306.2573 [hep-ph].
  %%CITATION = ARXIV:1306.2573;%%
  J.~Ellis, V.~Sanz and T.~You,
  %``Associated Production Evidence against Higgs Impostors and Anomalous Couplings,''
  arXiv:1303.0208 [hep-ph].
  %%CITATION = ARXIV:1303.0208;%%
  %6 citations counted in INSPIRE as of 28 Jun 2013
   M.~Muhlleitner, R.~M.~Godbole, C.~Hangst, S.~D.~Rindani and P.~Sharma,
  %``Analysis of Higgs spin and CP properties in a model-independent way in e+ e- ---> t anti-t Phi,''
  Frascati Phys.\ Ser.\  {\bf 54} (2012) 188.
  %%CITATION = 00309,54,188;%%
  
\bibitem{NSUSY}
J.~R.~Espinosa, C.~Grojean, V.~Sanz and M.~Trott,
  %``NSUSY fits,''
  arXiv:1207.7355 [hep-ph].
  %%CITATION = ARXIV:1207.7355;%%

\bibitem{Sally-NWA}
 A.~Djouadi,
  %``The Anatomy of electro-weak symmetry breaking. I: The Higgs boson in the standard model,''
  Phys.\ Rept.\  {\bf 457}, 1 (2008)
  [hep-ph/0503172].
  %%CITATION = HEP-PH/0503172;%%

 \bibitem{Paco}
 F.~del Aguila, M.~Perez-Victoria and J.~Santiago,
  %``Effective description of quark mixing,''
  Phys.\ Lett.\ B {\bf 492}, 98 (2000)
  [hep-ph/0007160].
   F.~del Aguila, M.~Perez-Victoria and J.~Santiago,
  %``Observable contributions of new exotic quarks to quark mixing,''
  JHEP {\bf 0009}, 011 (2000)
  [hep-ph/0007316].
   F.~del Aguila, J.~de Blas and M.~Perez-Victoria,
  %``Effects of new leptons in Electroweak Precision Data,''
  Phys.\ Rev.\ D {\bf 78}, 013010 (2008)
  [arXiv:0803.4008 [hep-ph]].
   F.~del Aguila, J.~de Blas and M.~Perez-Victoria,
  %``Electroweak Limits on General New Vector Bosons,''
  JHEP {\bf 1009}, 033 (2010)
  [arXiv:1005.3998 [hep-ph]].
  
  \bibitem{HagiwaraPRD}
    K.~Hagiwara, S.~Ishihara, R.~Szalapski and D.~Zeppenfeld,
  %``Low-energy effects of new interactions in the electroweak boson sector,''
  Phys.\ Rev.\ D {\bf 48} (1993) 2182.
  
  \bibitem{DeRujula}
 A.~De Rujula, M.~B.~Gavela, P.~Hernandez and E.~Masso,
  %``The Selfcouplings of vector bosons: Does LEP-1 obviate LEP-2?,''
  Nucl.\ Phys.\ B {\bf 384} (1992) 3.
  %%CITATION = NUPHA,B384,3;%%
  
  %\cite{Hagiwara:1992eh}
\bibitem{Hagiwara:1992eh}
  K.~Hagiwara, S.~Ishihara, R.~Szalapski and D.~Zeppenfeld,
  %``Low-energy constraints on electroweak three gauge boson couplings,''
  Phys.\ Lett.\ B {\bf 283} (1992) 353.
   O.~J.~P.~Eboli, M.~C.~Gonzalez-Garcia, S.~M. Lietti and S.~F.~Novaes,
  %``Probing intermediate mass Higgs interactions at the CERN Large Hadron Collider,''
  Phys.\ Lett.\ B {\bf 478} (2000) 199
  [hep-ph/0001030].
  M.~C.~Gonzalez-Garcia,
  %``Anomalous Higgs couplings,''
  Int.\ J.\ Mod.\ Phys.\ A {\bf 14} (1999) 3121
  [hep-ph/9902321].
   F.~Bonnet, M.~B.~Gavela, T.~Ota and W.~Winter,
  %``Anomalous Higgs couplings at the LHC, and their theoretical interpretation,''
  Phys.\ Rev.\ D {\bf 85} (2012) 035016
  [arXiv:1105.5140 [hep-ph]].
   S.~S.~Biswal, R.~M.~Godbole, B.~Mellado and S.~Raychaudhuri,
  %``Azimuthal Angle Probe of Anomalous $HWW$ Couplings at a High Energy $ep$ Collider,''
  Phys.\ Rev.\ Lett.\  {\bf 109} (2012) 261801
  [arXiv:1203.6285 [hep-ph]].
  %%CITATION = ARXIV:1203.6285;%%
  %5 citations counted in INSPIRE as of 28 Jun 2013
  
  
  \bibitem{Corbett:2012dm}
  T.~Corbett, {\it et. al} in \cite{fits}.
  
  \bibitem{SILH} 
  G.~F.~Giudice, C.~Grojean, A.~Pomarol and R.~Rattazzi,
  %``The Strongly-Interacting Light Higgs,''
  JHEP {\bf 0706}, 045 (2007)
  [hep-ph/0703164].

  %\cite{ATLAS:2012mec}
  \bibitem{Bonnet}
 F.~Bonnet, T.~Ota, M.~Rauch and W.~Winter,
  %``Interpretation of precision tests in the Higgs sector in terms of physics beyond the Standard Model,''
  arXiv:1207.4599 [hep-ph].
  %%CITATION = ARXIV:1207.4599;%%
  F.~Bonnet, M.~B.~Gavela, T.~Ota and W.~Winter,
  %``Anomalous Higgs couplings at the LHC, and their theoretical interpretation,''
  Phys.\ Rev.\ D {\bf 85}, 035016 (2012)
  [arXiv:1105.5140 [hep-ph]].
  %%CITATION = ARXIV:1105.5140;%%  
  
    \bibitem{PDG}
  J. Beringer et al. (Particle Data Group), Phys. Rev. D86,  (2012) 010001.
  
  %\cite{Alam:1997nk}
\bibitem{Alam:1997nk}
  S.~Alam, S.~Dawson and R.~Szalapski,
  %``Low-energy constraints on new physics revisited,''
  Phys.\ Rev.\ D {\bf 57} (1998) 1577
  [hep-ph/9706542].
  %%CITATION = HEP-PH/9706542;%%
  


%\cite{Erler:2012wz}
\bibitem{Erler:2012wz} 
  J.~Erler,
  %``Tests of the Electroweak Standard Model,''
  arXiv:1209.3324 [hep-ph].

%\cite{Peskin:1991sw}
\bibitem{Peskin:1991sw} 
  M.~E.~Peskin and T.~Takeuchi,
  %``Estimation of oblique electroweak corrections,''
  Phys.\ Rev.\ D {\bf 46}, 381 (1992).
  %%CITATION = PHRVA,D46,381;%%

\bibitem{ATLAS:2012mec} 
  G.~Aad {\it et al.}  [ATLAS Collaboration],
  %``Measurement of $W^+\W^-$ production in $pp$ collisions at $\sqrt{s}=7$ TeV with the ATLAS detector and limits on anomalous $WWZ$ and $WW\gamma$ couplings,''
  arXiv:1210.2979 [hep-ex].
  
\bibitem{Erler} 
  J.~Erler, private communication.

  \bibitem{LEP-direct}
    P.~Achard {\it et al.}  [L3 Collaboration],
  %``Search for anomalous couplings in the Higgs sector at LEP,''
  Phys.\ Lett.\ B {\bf 589}, 89 (2004)
  [hep-ex/0403037].
  %%CITATION = HEP-EX/0403037;%%
  
\bibitem{ATLASAA}
 G.~Aad {\it et al.}  [ATLAS Collaboration],
  %``Search for the Standard Model Higgs boson in the diphoton decay channel with 4.9 fb-1 of pp collisions at sqrt(s)=7 TeV with ATLAS,''
  Phys.\ Rev.\ Lett.\  {\bf 108}, 111803 (2012)
  [arXiv:1202.1414 [hep-ex]].
  %%CITATION = ARXIV:1202.1414;%%
  
\bibitem{CMSAA}
 S.~Chatrchyan {\it et al.}  [CMS Collaboration],
  %``Search for the standard model Higgs boson decaying into two photons in pp collisions at sqrt(s)=7 TeV,''
  Phys.\ Lett.\ B {\bf 710}, 403 (2012)
  [arXiv:1202.1487 [hep-ex]].
  %%CITATION = ARXIV:1202.1487;%%
  
 \bibitem{Feynrules}
 N.~D.~Christensen and C.~Duhr,
 {\it FeynRules - Feynman rules made easy,}
  Comput.\ Phys.\ Commun.\  {\bf 180} (2009) 1614
  [arXiv:0806.4194 [hep-ph]].
  %%CITATION = ARXIV:0806.4194;%%
  

\bibitem{MG5} 
  J.~Alwall {\it et al.}
{\it MadGraph 5 : Going Beyond,}
  JHEP {\bf 1106}, 128 (2011)
  [arXiv:1106.0522 [hep-ph]].
  %%CITATION = ARXIV:1106.0522;%%
  
  \bibitem{UFO}
 C.~Degrande, C.~Duhr, B.~Fuks, D.~Grellscheid, O.~Mattelaer and T.~Reiter,
{\it UFO - The Universal FeynRules Output,}
  Comput.\ Phys.\ Commun.\  {\bf 183} (2012) 1201
  [arXiv:1108.2040 [hep-ph]].
  %%CITATION = ARXIV:1108.2040;%%
  
\bibitem{PYTHIA} 
  T.~Sjostrand, S.~Mrenna and P.~Z.~Skands,
  {\it PYTHIA 6.4 Physics and Manual,}
  JHEP {\bf 0605}, 026 (2006)
  [hep-ph/0603175].
  %%CITATION = HEP-PH/0603175;%%
  
\bibitem{Delphes} 
  S.~Ovyn, X.~Rouby and V.~Lemaitre,
  {\it DELPHES, a framework for fast simulation of a generic collider experiment,}
  arXiv:0903.2225 [hep-ph].
  %%CITATION = ARXIV:0903.2225;%%
 




\bibitem{DaeSung}
 J.~Ellis and D.~S.~Hwang,
  %``Does the `Higgs' have Spin Zero?,''
  JHEP {\bf 1209}, 071 (2012)
  [arXiv:1202.6660 [hep-ph]].
  %%CITATION = ARXIV:1202.6660;%%
  
\bibitem{TaoHan}
 T.~Han, Y.~-P.~Kuang and B.~Zhang,
  %``Anomalous gauge couplings of the Higgs boson at high energy photon colliders,''
  Phys.\ Rev.\ D {\bf 73}, 055010 (2006)
  [hep-ph/0512193].
  %%CITATION = HEP-PH/0512193;%%
  

  
  \bibitem{Takubo}
   Y.~Takubo, R.~N.~Hodgkinson, K.~Ikematsu, K.~Fujii, N.~Okada and H.~Yamamoto,
  %``Measuring Anomalous Couplings in H->WW* Decays at the International Linear Collider,''
  arXiv:1011.5805 [hep-ph].
  %%CITATION = ARXIV:1011.5805;%%
  
 \bibitem{Hankele}
  V.~Hankele, G.~Klamke, D.~Zeppenfeld and T.~Figy,
  %``Anomalous Higgs boson couplings in vector boson fusion at the CERN LHC,''
  Phys.\ Rev.\ D {\bf 74}, 095001 (2006)
  [hep-ph/0609075].
  %%CITATION = HEP-PH/0609075;%%
  
 \bibitem{Plehn}
  T.~Plehn, D.~L.~Rainwater and D.~Zeppenfeld,
  %``Determining the structure of Higgs couplings at the LHC,''
  Phys.\ Rev.\ Lett.\  {\bf 88}, 051801 (2002)
  [hep-ph/0105325].
  %%CITATION = HEP-PH/0105325;%%
    
 \bibitem{Stolarski} 
   D.~Stolarski and R.~Vega-Morales,
  %``Directly Measuring the Tensor Structure of the Scalar Coupling to Gauge Bosons,''
  arXiv:1208.4840 [hep-ph].
  %%CITATION = ARXIV:1208.4840;%%
  
  \bibitem{ATLASWW}
   G.~Aad {\it et al.}  [ATLAS Collaboration],
  %``Search for the Standard Model Higgs boson in the H -> WW(*) -> l nu l nu decay mode with 4.7 /fb of ATLAS data at sqrt(s) = 7 TeV,''
  Phys.\ Lett.\ B {\bf 716}, 62 (2012)
  [arXiv:1206.0756 [hep-ex]].
  %%CITATION = ARXIV:1206.0756;%%
  
  \bibitem{CMSWW}
 S.~Chatrchyan {\it et al.}  [CMS Collaboration],
  %``Search for the standard model Higgs boson decaying to a W pair in the fully leptonic final state in pp collisions at sqrt(s) = 7 TeV,''
  Phys.\ Lett.\ B {\bf 710}, 91 (2012)
  [arXiv:1202.1489 [hep-ex]].
  %%CITATION = ARXIV:1202.1489;%%
  
\bibitem{ATLASZZ}
 G.~Aad {\it et al.}  [ATLAS Collaboration],
  %``Search for the Standard Model Higgs boson in the decay channel H->ZZ(*)->4l with 4.8 fb-1 of pp collision data at sqrt(s) = 7 TeV with ATLAS,''
  Phys.\ Lett.\ B {\bf 710}, 383 (2012)
  [arXiv:1202.1415 [hep-ex]].
  %%CITATION = ARXIV:1202.1415;%%
  
  \bibitem{CMSZZ}
   S.~Chatrchyan {\it et al.}  [CMS Collaboration],
  %``Search for the standard model Higgs boson in the decay channel H to ZZ to 4 leptons in pp collisions at sqrt(s) = 7 TeV,''
  Phys.\ Rev.\ Lett.\  {\bf 108}, 111804 (2012)
  [arXiv:1202.1997 [hep-ex]].
  %%CITATION = ARXIV:1202.1997;%%
  
  \bibitem{Ian-photonZ}
 J.~S.~Gainer, W.~-Y.~Keung, I.~Low and P.~Schwaller,
  %``Looking for a light Higgs boson in the $Z \gamma \to \ell \ell \gamma$ channel,''
  Phys.\ Rev.\ D {\bf 86}, 033010 (2012)
  [arXiv:1112.1405 [hep-ph]].
  %%CITATION = ARXIV:1112.1405;%%
  
\bibitem{revisited}
 A.~Djouadi, V.~Driesen, W.~Hollik and A.~Kraft,
  %``The Higgs photon - Z boson coupling revisited,''
  Eur.\ Phys.\ J.\ C {\bf 1}, 163 (1998)
  [hep-ph/9701342].
  %%CITATION = HEP-PH/9701342;%%

\bibitem{haa}
 A.~Djouadi, V.~Driesen, W.~Hollik and J.~I.~Illana,
  %``The Coupling of the lightest SUSY Higgs boson to two photons in the decoupling regime,''
  Eur.\ Phys.\ J.\ C {\bf 1} (1998) 149
  [hep-ph/9612362].
  %%CITATION = HEP-PH/9612362;%%
  
\bibitem{Csaba-Jay}
 C.~Csaki, J.~Hubisz and S.~J.~Lee,
  %``Radion phenomenology in realistic warped space models,''
  Phys.\ Rev.\ D {\bf 76}, 125015 (2007)
  [arXiv:0705.3844 [hep-ph]].
  %%CITATION = ARXIV:0705.3844;%%

\bibitem{little-RS}
 H.~Davoudiasl, G.~Perez and A.~Soni,
  %``The Little Randall-Sundrum Model at the Large Hadron Collider,''
  Phys.\ Lett.\ B {\bf 665}, 67 (2008)
  [arXiv:0802.0203 [hep-ph]].
  %%CITATION = ARXIV:0802.0203;%%


\bibitem{pdg}
 S.~Eidelman {\it et al.}  [Particle Data Group Collaboration],
  %``Review of particle physics. Particle Data Group,''
  Phys.\ Lett.\ B {\bf 592} (2004) 1.
  %%CITATION = PHLTA,B592,1;%%


\end{thebibliography}
\end{document}